\documentclass[reprint,showpacs,superscriptaddress,english,twocolumn,notitlepage,longbibliography,nofootinbib]{revtex4-1}
\usepackage{times}
\usepackage{graphicx}
\usepackage{float}
\usepackage{latexsym,amsmath,amssymb,bm,euscript}
\usepackage{color}
\usepackage{subfigure}
\usepackage{epstopdf}
\usepackage[colorlinks=true,linkcolor=blue,citecolor=blue,urlcolor=blue]{hyperref}
\usepackage{hyperref}
\usepackage{ulem}
\usepackage{cleveref}
\usepackage{tikz}
\usetikzlibrary{decorations.pathreplacing}

\usepackage{ytableau}

\definecolor{colorhhy}{rgb}{0.9, 0.17, 0.31}

\crefname{appendix}{Appendix}{Appendices}
\crefname{equation}{Eq.}{Eqs.}
\crefname{figure}{Fig.}{Figs.}
\crefname{table}{Table}{Tables}
\crefname{section}{Appendix}{Appendices}
\crefname{enumi}{Point}{Points}
\creflabelformat{appendix}{[#2#1#3]}

\crefmultiformat{figure}{Figs.~#2#1\xdef\mycreffirstarg{#1}#3}%
 { and~#2#1#3}{, #2#1#3}{ and~#2#1#3}
\newcommand{\be}[0]{\begin{equation}}
\newcommand{\ee}[0]{\end{equation}}

\def\ba#1\ea{\begin{align}#1\end{align}}

\newcommand{\bmat}[0]{\begin{bmatrix}}
\newcommand{\emat}[0]{\end{bmatrix}}

\begin{document}

\title{Two-dimensional moir\'{e} phonon polaritons}

\author{Hao Shi}
\affiliation{New Cornerstone Science Laboratory, Department of Physics, The Hong Kong University of Science and Technology, Hong Kong, China}
\affiliation{Department of Physics, The Hong Kong University of Science and Technology, Hong Kong, China}
\author{Chu Li}
\affiliation{Department of Physics, The Hong Kong University of Science and Technology, Hong Kong, China}
\author{Ding Pan}
\affiliation{Department of Physics, The Hong Kong University of Science and Technology, Hong Kong, China}
\affiliation{Department of Chemistry, The Hong Kong University of Science and Technology, Hong Kong, China}
\author{Xi Dai}
\affiliation{New Cornerstone Science Laboratory, Department of Physics, The Hong Kong University of Science and Technology, Hong Kong, China}
\affiliation{Department of Physics, The Hong Kong University of Science and Technology, Hong Kong, China}

\begin{abstract} \label{abstract}
Phonon polaritons (PhPs) are hybrid light-matter modes. We investigate them in two-dimensional (2D) materials with twisted moir\'{e} structures, revealing that the moir\'{e} potential creates a new class of `moir\'{e} PhPs'. These exhibit a fundamental spectral reconstruction into multiple branches and, crucially, electromagnetic wavefunctions that are nano-patterned by the superlattice. Through numerical simulations based on realistic lattice models, we confirm the existence of these intriguing modes. The inherent nanoscale structuring produces a robust, spatially varying near-field response, establishing moir\'{e} superlattices as a platform for engineering light-matter interactions.
\end{abstract}

\maketitle

{\it Introduction~~}
Polaritons arise from the coupling of photons with collective excitations in materials, such as phonons, plasmons, and excitons. These hybrid modes exhibit properties of both light and matter, enabling broad applications in fields like optics \cite{21n_optics_polariton,20nc_phonon_laser}, condensed matter physics \cite{22np_zeeman_polariton,16s_2D_polariton}, and quantum computing \cite{20npjqi_polariton_QC,22nrp_polariton_QC}. In polar crystals, ions oscillate with polarization and interact with electromagnetic (EM) waves. The coupling between ionic motion and the EM field produces phonon polariton (PhPs). The first PhP model for 3D crystals was established by Huang's equations \cite{51n_PhP,96oxford_book}, which treat long-wavelength ionic vibrations and polarization macroscopically. Solving Huang's equations alongside Maxwell's equations yields 3D PhPs. A similar macroscopic theory can also be applied to 2D materials, though it incorporates additional constraints from EM boundary conditions \cite{17_polariton_review,23s_moire_optics_review,17nl_LOTO_degeneracy,19nl_2D_polariton,20acsp_acoustic_PhP,21nl_2D_phonon_oxide,21nm_BN_polariton_exp,24nc_nonanalytic_BN_exp,83pssb_heterobilayer}. In 2D systems, PhP can manifest as transverse magnetic (TM) or transverse electric (TE) modes, propagating along the material surface.

Moir\'{e} superlattices offer a novel approach to engineer 2D physics at length scales far exceeding the crystal periodicity, serving as a powerful platform for light-matter interactions \cite{16s_2D_polariton,24prb_plasmon_inhomo_ttg,24prb_acoustic_plasmon_tbg,20n_twist_phonon_exp}. 
The discovery of superconducting and correlated insulating states in twisted bilayer graphene \cite{18n_tbg_sc_cao,18n_ci_tbg} has spurred the observation of exotic phenomena in moir\'{e} systems \cite{21n_sc_TTG,20n_ci_moire,24n_ci_RMG,20s_QAH,23n_FQAH_cai,24n_FQAH_kang,13njp_tbg_plasmon,16nl_tbg_plasmon_calculation,21prb_tbg_plasmon_cal_TB,24prb_plasmon_inhomo_ttg,24prb_acoustic_plasmon_tbg,23sa_plasmon_correlation,21n_exciton_polariton_exp,22nl_exciton_twist_tuning,24a_exciton_polariton_electric}. Despite widespread interest and progress, PhPs in moir\'{e} systems remain poorly explored, likely due to the limited optical resolution of the tiny energy scales characteristic of moir\'{e} physics. Previous work has explored PhPs primarily in thicker twisted structures where modulation of polariton propagation dominates \cite{23nm_Dai_review,20nl_Zheng,20nm_chen,20n_Hu_PhP,23nm_duan,21nc_twisted_hBN_exp,20nc_PhP_graphene_hBN}. However, study focused on atomically thin layers is missing. Additionally, experimental samples often exhibit high dissipation, complicating direct detection of moir\'{e} polaritons. Theoretically, the challenge lies in managing the vast degrees of freedom inherent to moir\'{e} superlattices.

In this study, we investigate PhPs in moir\'{e} materials-specifically, twisted bilayer hexagonal boron nitride (hBN) and MoTe$_2$, using lattice models. We reveal that the moir\'{e} potential gives rise to a new class of PhPs with two defining characteristics: (I) a fundamental spectral reconstruction into multiple, flat PhP branches (Fig. \ref{fig:fig_PhP_dispersion}), and (II) most importantly, electromagnetic wavefunctions that are nano-patterned by the moir\'{e} lattice itself. This results in a unique physical phenomenon: long-wavelength evanescent light can excite confined optical states with spatial features orders of magnitude smaller than the photon's wavelength—a form of inherent nanoscale optical structuring absent in conventional materials. This manifests as a spatially inhomogeneous local response \cite{24prb_plasmon_inhomo_ttg,24prb_acoustic_plasmon_tbg} that provides a robust, experimentally accessible signature via near-field techniques (Fig. \ref{fig:fig_local_Pi}) \cite{21nc_twisted_hBN_exp}, even when the fine spectral details are obscured by a realistic phonon linewidth. Thus, the moiré potential provides a powerful new knob to actively engineer light-matter interactions at the nanoscale.

\begin{figure}
\includegraphics[width=0.43\textwidth]{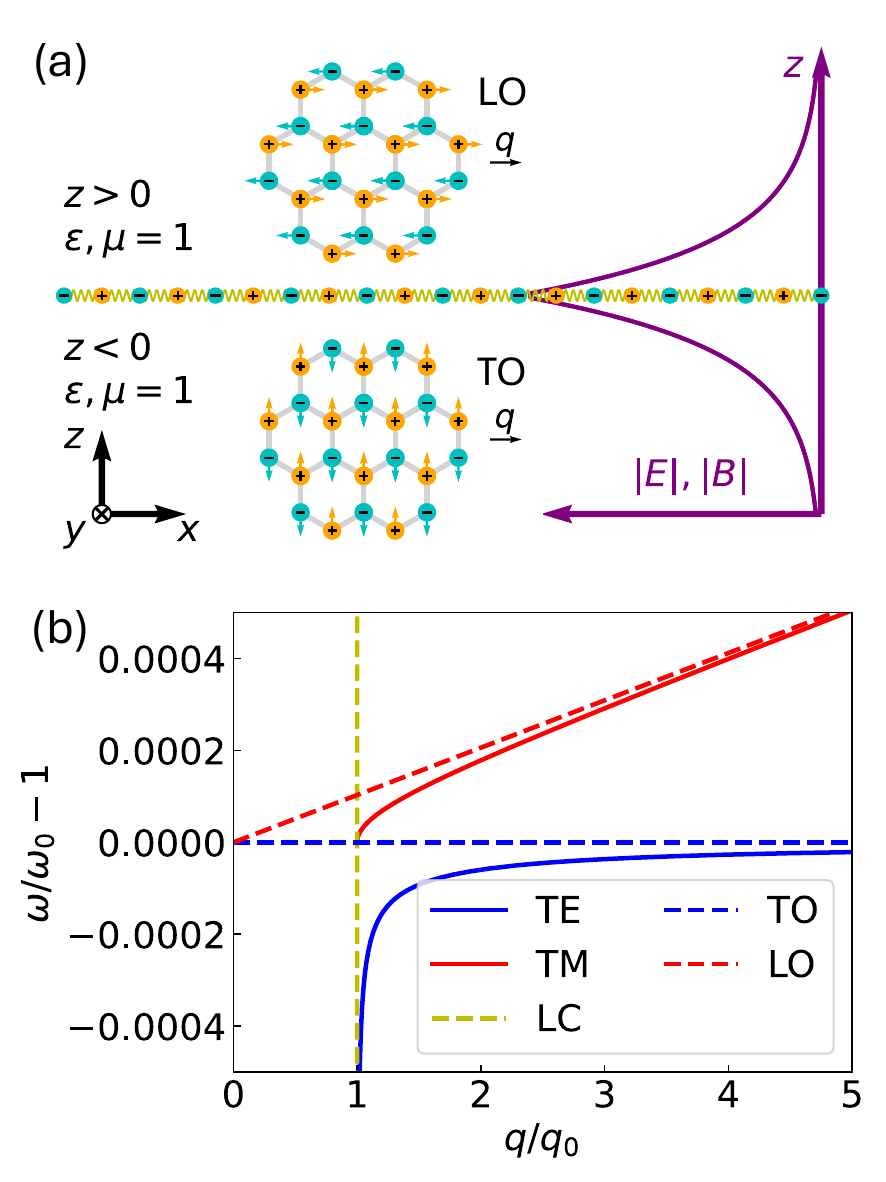}
\caption{\label{fig:fig_basic}(a) A 2D polar sheet is positioned at $z=0$ in vacuum. The PhP exhibits characteristic 2D EM waves that decay along the $z$-axis, as illustrated by the purple coordinate system. The inset displays the long-wavelength ($\bm{q}=q\bm{e}_x$) LO and TO modes patterns in the $xy$-plane for a binary crystal. (b) The 2D PhP dispersion of the TM, TE modes near the light cone (LC) and resonance frequency $\omega_0$. For comparison, the LO, TO modes under the non-retarded approximation are also shown. In (b), we use $T/(2\omega_0c)=2.06\times10^{-4}$, obtained from the lattice model of monolayer hBN (SI Section 2.2).}
\end{figure}

{\it Theoretical formalism}~~
We begin with the 2D PhP formalism. Consider an ionic sheet positioned at $z=0$ in vacuum [$\epsilon,\mu=1$, Fig. \ref{fig:fig_basic}(a)]. Its dynamics are governed by the vibration field $\bm{W}$ describing the in-plane ionic motion, which obeys the equation of motion:
\begin{align}
\ddot{\bm{W}}=-\omega_0^2\bm{W}+\gamma_{12}\bm{E}_t, \label{eq:EoM}
\end{align}
where $\omega_0$ is the resonance frequency, and $\bm{E}_t$ denotes the in-plane component of the electric field $\bm{E}$ at $z=0$. The in-plane polarization density $\bm{P}$ arises primarily from ionic displacement, 
\begin{align}
\bm{P}=\gamma_{21}\bm{W}. \label{eq:P_WE}
\end{align}
Here, $\gamma_{12}=\gamma_{21}=\sqrt{\varepsilon_0T}$ can be derived from microscopic models. These equations represent the 2D analogs of Huang's equations and must be solved together with Maxwell's equations and the boundary conditions at $z=0$. We seek solutions of the form $\bm{E},\bm{W}\propto e^{i\bm{q}\cdot\bm{r}-i\omega t}$, where $\bm{r}$ and $\bm{q}$ are the \textit{in-plane} position and momentum, respectively. The susceptibility is then defined as
\begin{align}
\bm{P}=\varepsilon_0\Pi(\omega)\bm{E}_t,\quad\Pi(\omega)=\frac{T}{\omega_0^2-\omega^2}. \label{eq:toy_polarization}
\end{align}
The above equations have guided or radiative solutions \cite{66pr_slab_PhP_guided,66pr_slab_PhP_radiative}, depending on whether the decaying parameter $\lambda=\sqrt{q^2-\omega^2/c^2}$ is real or imaginary. Radiative modes correspond to conventional light propagation problems with the polar sheet acting as a scattering interface [Supporting Information (SI) Section 1.4]. Our focus, however, is on guided modes that feature localized 2D EM waves near $z=0$ \cite{89spj_retardation_effect}. The guided modes split into an $s$-polarized (TE) mode with $\bm{E}\perp\bm{q}$ and a $p$-polarized (TM) mode with $\bm{B}\perp\bm{q}$. The dispersions of the TE and TM modes are shown in Fig. \ref{fig:fig_basic}(b), which are governed by the eigen equations $1-\Pi(\omega)\omega^2/(2\lambda c^2)=0$ and $1+\lambda\Pi(\omega)/2=0$, respectively (SI Section 1.3). The TE mode resembles free-space light at $q\ll\omega_0/c$, while it converges to pure lattice oscillations at $q\gg\omega_0/c$. The TM mode's (long-wavelength) dispersion starts at $\omega_0=cq_0$ and tends to linear dispersion at $q\gg\omega_0/c$. They are quintessential 2D EM waves with a power density localized along $z$, arising universally in 2D materials and 3D material interfaces due to polarizable collective modes. The conditions to determine the eigenmodes are quite general: e.g., substituting $\Pi(\omega)$ with its plasmonic counterpart extends the framework to 2D plasmon polaritons. Critically, TM (TE) mode requires $\Pi(\omega)<0$ [$\Pi(\omega)>0$]. This sign rule for polarization persists-for instance, graphene's interband conductivity enables $\Pi(\omega)<0$ in a specific regime, hosting a unique TE plasmon mode \cite{07prl_TE_graphene,16sr_TE_graphene_exp} absent in conventional 2D electron gas \cite{67prl_2DEG_plasmon}.

It is instructive to consider the non-retarded limit ($q \gg \omega/c$), where retardation effects are neglected and the Coulomb interaction is treated as instantaneous. In this limit, the TM and TE modes reduce to the transverse optical (TO) and longitudinal optical (LO) phonon modes, respectively (SI Section 1.2). Their dispersions are shown as dashed lines in Fig. \ref{fig:fig_basic}(b). The TO mode corresponds to a pure mechanical oscillation where $\bm{E}=\bm{0}$ and $\bm{W}\perp\bm{q}$ vibrates at a fixed frequency $\omega_{\text{TO}}=\omega_0$. In contrast, the LO mode involves a macroscopic $\bm{E}$ field that couples to the vibration. Its dispersion is governed by $1 + q\Pi(\omega)/2 = 0$. From this, a characteristic linear LO-TO splitting can be derived in the long-wavelength limit: $\omega_{\text{LO}}-\omega_\text{TO}\approx qT/(4\omega_0)$. This linear splitting is a fundamental signature of 2D polar systems \cite{17nl_LOTO_degeneracy,19nl_2D_polariton,21nl_2D_phonon_oxide,24nc_nonanalytic_BN_exp}, arising from the long-range Coulomb interaction in a reduced dimension. It stands in stark contrast to the behavior in 3D bulk crystals, where the large depolarizing field leads to a $q$-independent splitting at the Brillouin zone center \cite{51n_PhP,96oxford_book}. This key difference highlights the profound impact of dimensionality on light-matter interactions in polar materials.

Both guided and radiative modes can also be treated within a unified framework of light reflection and refraction (SI Section 1.5). In this approach, the PhP dispersion $\omega(\bm{q})$ emerges as the poles of the transmission matrix $T(\bm{q},\omega)$, offering computational advantages \cite{16ws_book_plasmon}. The spectrum can be visualized by plotting $\mathcal{L}(\bm{q},\omega)=-\text{Im}[\text{det}[T(\bm{q},\omega+i\delta/2)]]$, where $\delta$ (representing the phonon linewidth) is tiny. This method simultaneously captures the continuous spectrum of radiative solutions and the discrete dispersions of guided modes.

\begin{figure*}
\centering
\includegraphics[width=1.0\textwidth]{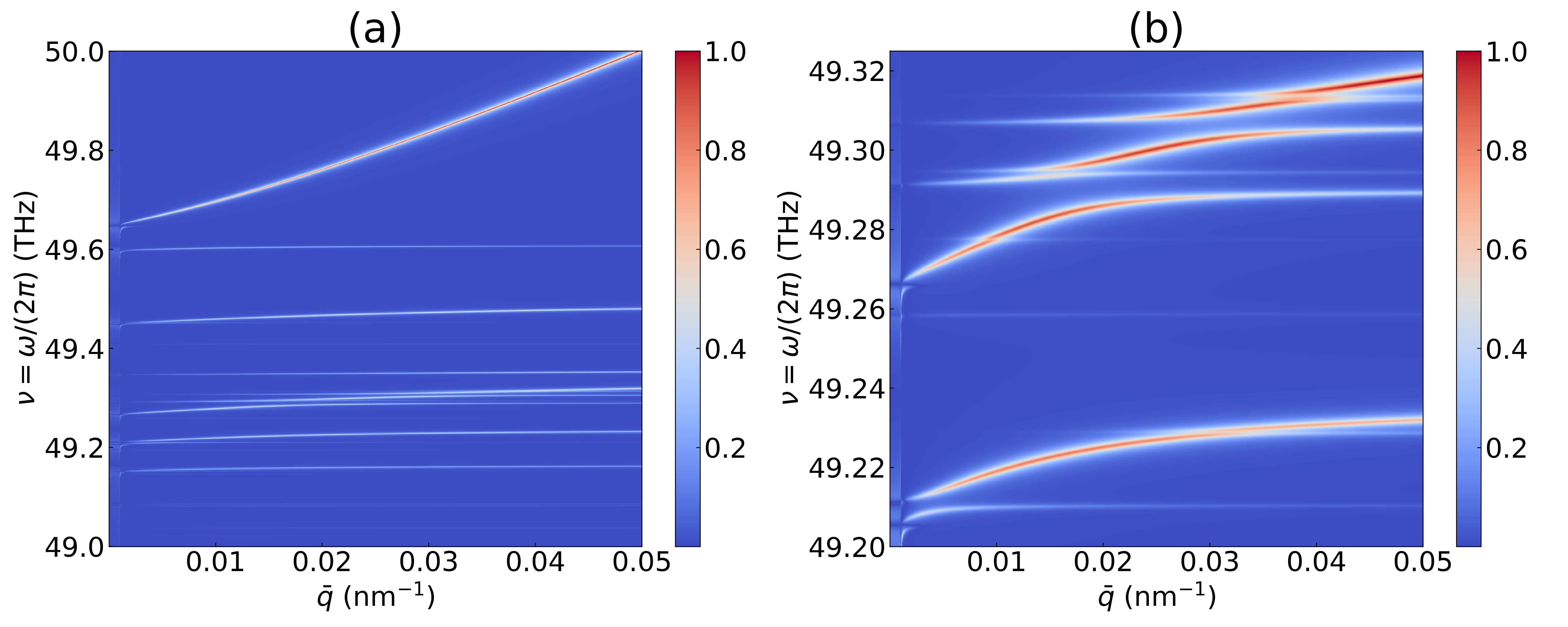}
\caption{\label{fig:fig_PhP_dispersion} (a) The long-wavelength PhP dispersion of $2.65^{\circ}$ twisted bilayer hBN near $\nu_0=\omega_0/(2\pi)\approx 49.5$ THz ($q_0\approx 10^{-3}$ nm$^{-1}$), along the $\bar{\Gamma}-\bar{M}$ line, obtained by plotting the (normalized) spectrum $\ln(1+|\mathcal{L}(\bar{\bm{q}},\omega)|)$. Here a tiny linewidth $\delta/(2\pi)=10^{-3}$ THz is used to make each branch distinguishable. Many flat branches appear below the topmost dominant branch. (b) The detailed dispersion within the mini window $49.20$-$49.325$ THz.}
\end{figure*}

The physics becomes richer in moir\'{e} systems, where the supercell can reach mesoscopic scales with vast sublattice degrees of freedom \cite{21nc_twisted_hBN_exp}. Phonons folds into the moir\'{e} Brillouin zone (mBZ), generating intricate moir\'{e} phonon bands \cite{23prb_TAPW,25prb_moire_phonon_tbg}. This raises a compelling question: how do PhPs emerge in such complex systems amid long-range EM interactions? For quantitative analysis, we utilize realistic lattice models that bypass computationally intense \textit{ab initio} methods \cite{97prb_DFPT}. Short-range ionic interactions are modeled via a force field (SI Section 6), while long-range Coulomb forces are treated through macroscopic electric fields. The displacement $\bm{u}$ of an ion at position $\bm{r}_{Ii\alpha}$ (where $I$, $i$, and $\alpha$ index the supercell, atomic cell, and sublattice positions, respectively, as detailed in SI Section 3.1) satisfies the following equation of motion
\begin{align}
\begin{split}
M_{\alpha}\ddot{\bm{u}}(\bm{r}_{Ii\alpha})+\sum_{Jj\beta}\Phi_{i\alpha,j\beta}(\bm{r}_{Ii\alpha}-\bm{r}_{Jj\beta})\bm{u}(\bm{r}_{Jj\beta})\\
-\sum_{\bm{Q}}Z_{\alpha}e\bm{E}_{\bar{\bm{q}}+\bm{Q},t}e^{i(\bar{\bm{q}}+\bm{Q})\cdot\bm{r}_{Ii\alpha}-i\omega t}=0, \label{eq:EoM_moire}
\end{split}
\end{align}
where $\Phi$ is the force constant \cite{88prb_new_empirical,05prb_registry_graphitic,18nl_nanoserpents,19nl_mechanical,85prb_silicon,15nt_SWP_MoS2,21nl_anisotropic_interlayer, 23pca_anisotropic_interlayer, 22cpc_lammps}, $M_{\alpha}$ and $Z_{\alpha}$ are the ionic mass and charge (in units of $e$), respectively. The moir\'{e} electric field $\bm{E}_t=\sum_{\bm{Q}}\bm{E}_{\bar{\bm{q}}+\bm{Q},t}e^{i(\bar{\bm{q}}+\bm{Q})\cdot\bm{r}-i\omega t}$ includes components indexed by moir\'{e} reciprocal vectors $\bm{Q}$, with $\bar{\bm{q}}\in$mBZ. The final term in Eq. (\ref{eq:EoM_moire}) is the driving force from the macroscopic electric field, which encodes the long-range 2D Coulomb interaction essential for PhP formation. The polarization density is given by \cite{96oxford_book,18a_linear_response_Pi,19nl_2D_polariton,16prb_Frohlich_2D_reff}
\begin{align}
\bm{P}(\bm{r})=\sum_{Ii\alpha}Z_{\alpha}e\bm{u}(\bm{r}_{Ii\alpha})\delta(\bm{r}-\bm{r}_{Ii\alpha}). \label{eq:P_WE_moire}
\end{align}
These equations generalize Eqs. (\ref{eq:EoM}) and (\ref{eq:P_WE}) to the lattice level \cite{25a_Huang_lattice}. Without $\bm{E}_t$, Eq. (\ref{eq:EoM_moire}) reduces to the standard non-polar phonon problem. The driven harmonic oscillator system admits an exact solution \cite{96oxford_book}, yielding a susceptibility tensor with multiple poles due to the moir\'{e} potential (SI Section 3). In Fourier basis, $\bm{P}(\bm{r})=\sum_{\bm{Q}}\bm{P}_{\bar{\bm{q}}+\bm{Q}}e^{i(\bar{\bm{q}}+\bm{Q})\cdot\bm{r}-i\omega t}$, we obtain
\begin{align}
\begin{split}
&\bm{P}_{\bar{\bm{q}}+\bm{Q}}=\varepsilon_0\sum_{\bm{Q}'}\bm{\Pi}^{\bm{Q}\bm{Q}'}(\bar{\bm{q}},\omega)\bm{E}_{\bar{\bm{q}}+\bm{Q}',t},\\
&\bm{\Pi}^{\bm{Q}\bm{Q}'}(\bar{\bm{q}},\omega)=\frac{e^2}{\varepsilon_0\Omega_m}\sum_b\frac{\bm{S}_{\bm{Q}b}(\bar{\bm{q}})\bm{S}^{\dagger}_{\bm{Q}'b}(\bar{\bm{q}})}{\omega^2_{\bar{\bm{q}}b}-\omega^2}, \label{eq:Pi_result} 
\end{split}
\end{align}
where $\varepsilon_0$ is the vacuum permittivity, $\Omega_m$ is the supercell area, $\omega_{\bar{\bm{q}}b}$ and $\bm{e}_{b}(\bar{\bm{q}})$ the bare frequency and displacement vector of the $b$-th moir\'{e} phonon without $\bm{E}_t$, and the $S$ matrix is
\begin{align}
\bm{S}_{\bm{Q}b}(\bar{\bm{q}})=\sum_{i\alpha}\frac{Z_{\alpha}\bm{e}_{i\alpha,b}(\bar{\bm{q}})}{\sqrt{M_{\alpha}}}e^{-i\bm{Q}\cdot(\bm{R}_i+\bm{\tau}_{\alpha})}.\label{eq:Smat}
\end{align}
The moir\'{e} physics manifests in the off-diagonal terms of $\bm{\Pi}^{\bm{Q}\bm{Q}'}$. The $\bm{Q}\neq\bm{0}$ terms encode field modulations at moir\'{e} length scales \cite{24prb_plasmon_inhomo_ttg}. If we turn off the moir\'{e} potential, Eq. (\ref{eq:Pi_result}) becomes diagonal in $\bm{Q}$, recovering the moir\'{e}-free case (SI Section 5.4).

\begin{figure*}
\centering
\includegraphics[width=0.98\textwidth]{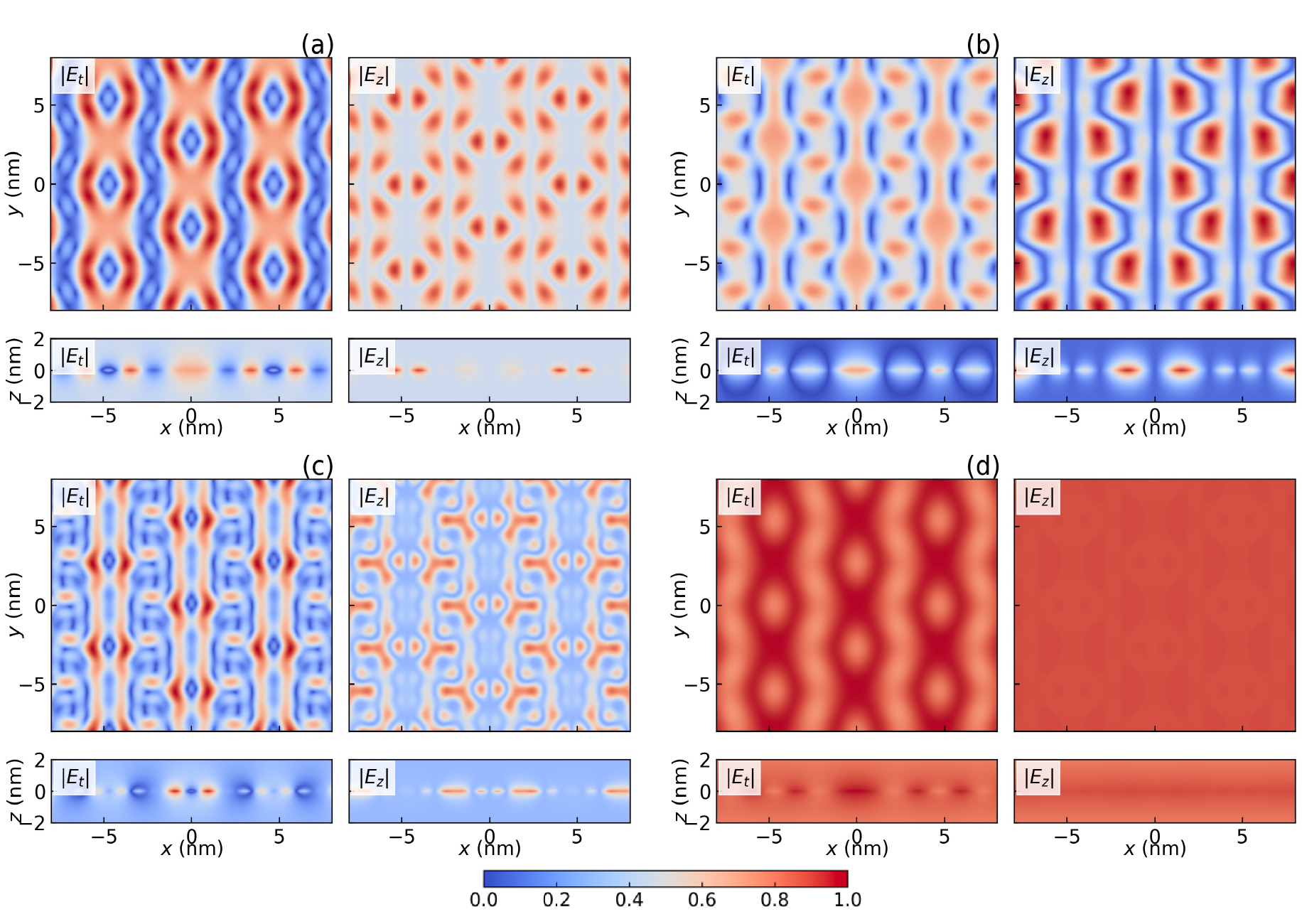}
\caption{\label{fig:fig_PhP_E} Field distributions of moir\'{e} PhPs in $2.65^{\circ}$ twisted bilayer hBN: in-plane ($z=0$ top row) and out-of-plane ($y=0$, bottom row) amplitudes $|\bm{E}_t|$ and $|E_z|$ along $\bar{\Gamma}-\bar{M}$ line at (a) $\bar{q}=0.01 \text{ nm}^{-1}$, $\nu=49.696$ THz; (b) $\bar{q}=0.01 \text{ nm}^{-1}$, $\nu=49.602$ THz; (c) $\bar{q}=0.01 \text{ nm}^{-1}$, $\nu=49.217$ THz; (d) $\bar{q}=0.05 \text{ nm}^{-1}$, $\nu=50.002$ THz. Fields are normalized to maxima of $1$. (a) and (c) indicate that, at a fixed $\bar{\bm{q}}$, the specific moir\'{e} pattern of EM waves is sensitive to the frequency. (a) and (d) are taken from the same branch.}
\end{figure*}

The moir\'{e} PhPs are determined by solving Maxwell's equations with appropriate boundary conditions. Assuming an infinitesimally thin moir\'{e} material for simplicity, the eigenmode problem reduces to solving the secular equation $\det[A(\bar{\bm{q}},\omega)]=0$, where $A(\bar{\bm{q}},\omega)$ is a block-structured matrix acting on the space of $\bm{Q}$, encoding the material's light-scattering properties (SI Section 3.2). The matrix elements are
\begin{align}
\begin{split}
&A^{\bm{Q}\bm{Q}'}_{\parallel\parallel}(\bar{\bm{q}},\omega)=\delta_{\bm{Q}\bm{Q}'}+\frac{\lambda_{\bar{\bm{q}}+\bm{Q}}}{2}\Pi^{\bm{Q}\bm{Q}'}_{\parallel\parallel}(\bar{\bm{q}},\omega),\\
&A^{\bm{Q}\bm{Q}'}_{\parallel\perp}(\bar{\bm{q}},\omega)= \frac{\lambda_{\bar{\bm{q}}+\bm{Q}}}{2}\Pi^{\bm{Q}\bm{Q}'}_{\parallel\perp}(\bar{\bm{q}},\omega),\\
&A^{\bm{Q}\bm{Q}'}_{\perp\parallel}(\bar{\bm{q}},\omega)=-\frac{1}{2\lambda_{\bar{\bm{q}}+\bm{Q}}}\frac{\omega^2}{c^2}\Pi^{\bm{Q}\bm{Q}'}_{\perp\parallel}(\bar{\bm{q}},\omega),\\
&A^{\bm{Q}\bm{Q}'}_{\perp\perp}(\bar{\bm{q}},\omega)=\delta_{\bm{Q}\bm{Q}'}-\frac{1}{2\lambda_{\bar{\bm{q}}+\bm{Q}}}\frac{\omega^2}{c^2}\Pi^{\bm{Q}\bm{Q}'}_{\perp\perp}(\bar{\bm{q}},\omega). \label{eq:Amat}
\end{split}
\end{align}
Here, $\parallel$ and $\perp$ denote components parallel and perpendicular to $\bar{\bm{q}}+\bm{Q}(\bm{Q}')$, respectively, with $\lambda_{\bar{\bm{q}}+\bm{Q}}=\sqrt{|\bar{\bm{q}}+\bm{Q}|^2-\omega^2/c^2}$. Equation (\ref{eq:Amat}) is the central result of our work, which contains all the information about moir\'{e} PhPs. The transmission matrix can be obtained from the $A$ matrix as: $T(\bar{\bm{q}},\omega)=A^{-1}(\bar{\bm{q}},\omega)$. The PhP dispersion can be obtained by searching the zeros of $\text{det}(A)$ [poles of $\det(T)$], and the corresponding eigenmodes can be obtained as the null vectors of $A$. A key feature of moir\'{e} PhPs is that an incident evanescent wave (with long in-plane wavelength) can excite EM fields with much shorter wavelengths. This occurs through moir\'{e} potential scattering, which is encoded in the off-diagonal elements (in $\bm{Q}$) of the scattering matrix $A(\bar{\bm{q}},\omega)$ (SI Section 3.2). So we focus exclusively on the case where the incident light has $\bm{Q}=\bm{0}$ components only. The effective transmission matrix is the long-wavelength block of the full transmission matrix $T_{\text{eff}}(\bar{\bm{q}},\omega)=[A^{-1}(\bar{\bm{q}},\omega)]^{\bm{0}\bm{0}}$ \cite{62pr_local_response,23sa_plasmon_correlation,24prb_acoustic_plasmon_tbg}.
The poles of the spectrum $\mathcal{L}(\bar{\bm{q}},\omega)=-\text{Im}[\text{det}[T_{\text{eff}}(\bar{\bm{q}},\omega+i\delta/2)]]$ depicts the dispersion of moir\'{e} PhPs that can be excited by long-wavelength light.

{\it Moir\'{e} PhP in hBN and MoTe$_2$}~~
We select hBN and MoTe$_2$ as two examples, which are popular insulating polar crystals \cite{21nc_twisted_hBN_exp,24a_tBN_exp,24a_tBN_exp2,23n_FQAH_cai,24n_FQAH_kang}. Our analysis focuses on AA-stacked twisted bilayer configurations of these materials. A different stacking style could slightly influence the PhP dispersion but would not alter the moir\'{e} physics discussed here. While our numerical examples focus on hexagonal lattices, the above formalism is general and applicable to any 2D moir\'{e} polar system.

Hexagonal boron nitride is a prototypical polar material for PhP studies \cite{14s_BN_exp_Dai,19nl_2D_polariton,20acsp_acoustic_PhP,21nl_2D_phonon_oxide,21nm_BN_polariton_exp,24nc_nonanalytic_BN_exp}, featuring an optical phonon frequency $\nu_0=\omega_0/(2\pi)\approx 49.4$ THz (calculated using a molecular dynamics-based lattice model to explore the qualitative physics of moir\'{e} PhPs; it is larger than the experimental value of $\sim41$ THz). We adopt isotropic charges $Z_\text{B}=-Z_\text{N}\approx 2.7$ \cite{21nl_2D_phonon_oxide} and focus on $2.65^{\circ}$ twisted bilayer hBN that has lattice length $L_{\theta}\approx 5.42$ nm and $1876$ atoms per supercell. The long-wavelength dispersion near $\omega_0$ is shown in Fig. \ref{fig:fig_PhP_dispersion}, where many PhP branches appear. Although the phonon moir\'{e} potential is weak in magnitude, it effectively hybridizes the long-wavelength ($\bm{Q}=\bm{0}$) components with shorter-wavelength ($\bm{Q}\neq\bm{0}$) components through non-negligible off-diagonal terms in the susceptibility tensor Eq. (\ref{eq:Pi_result}), particularly near the resonance frequency $\omega_{\bar{\bm{q}}b}$. This hybridization generates new PhP branches exhibiting characteristic moir\'{e} interference patterns. The resulting dispersions exhibit sharp transitions between spectral regions bounded by folded phonon frequencies $\omega_{\bar{\bm{q}}b}$, forming a series of mini-bands in the polariton spectrum. The dominant branch above $49.6$ THz resembles the TM mode without moir\'{e} potential. The neighboring phonon frequencies $\omega_{\bar{\bm{q}}b}$ stay very close to each other. Therefore, the emerging moir\'{e} modes are quite flat, with energy resolutions on the scale of $\sim$0.01 THz [Fig. \ref{fig:fig_PhP_dispersion}(b)]. This fine structure would be significantly obscured under a more realistic linewidth $\delta$ \cite{18prb_hBN_linewidth,18nm_hBN_linewidth,20prl_hBN_linewidth} (SI Section 3.3). Consequently, resolving the full moir\'{e} PhP dispersion poses a significant experimental challenge and requires samples with exceptionally low dissipation. All eigenmodes represent genuine moir\'{e} PhPs, as their EM fields (and lattice oscillations) exhibit varying degrees of wavelength mixing. The electric fields $\bm{E}$ for some representative modes are plotted in Fig. \ref{fig:fig_PhP_E}. Spatial modulations of $\bm{E}$ can be clearly seen in the $xy$-plane. Each PhP branch is characterized by a unique moir\'{e} pattern. Such patterns are sensitive to the branch frequency [Fig. \ref{fig:fig_PhP_E}(a) vs (b) vs (c)] and momentum $\bar{\bm{q}}$ [Fig. \ref{fig:fig_PhP_E}(a) vs (d)], while all the patterns are confined along the $z$-axis to a length scale $|\bm{Q}|^{-1}\sim L_{\theta}$. These characteristic spatial signatures of moir\'{e} PhPs are absent in moir\'{e}-free systems.

\begin{figure}
\includegraphics[width=0.46\textwidth]{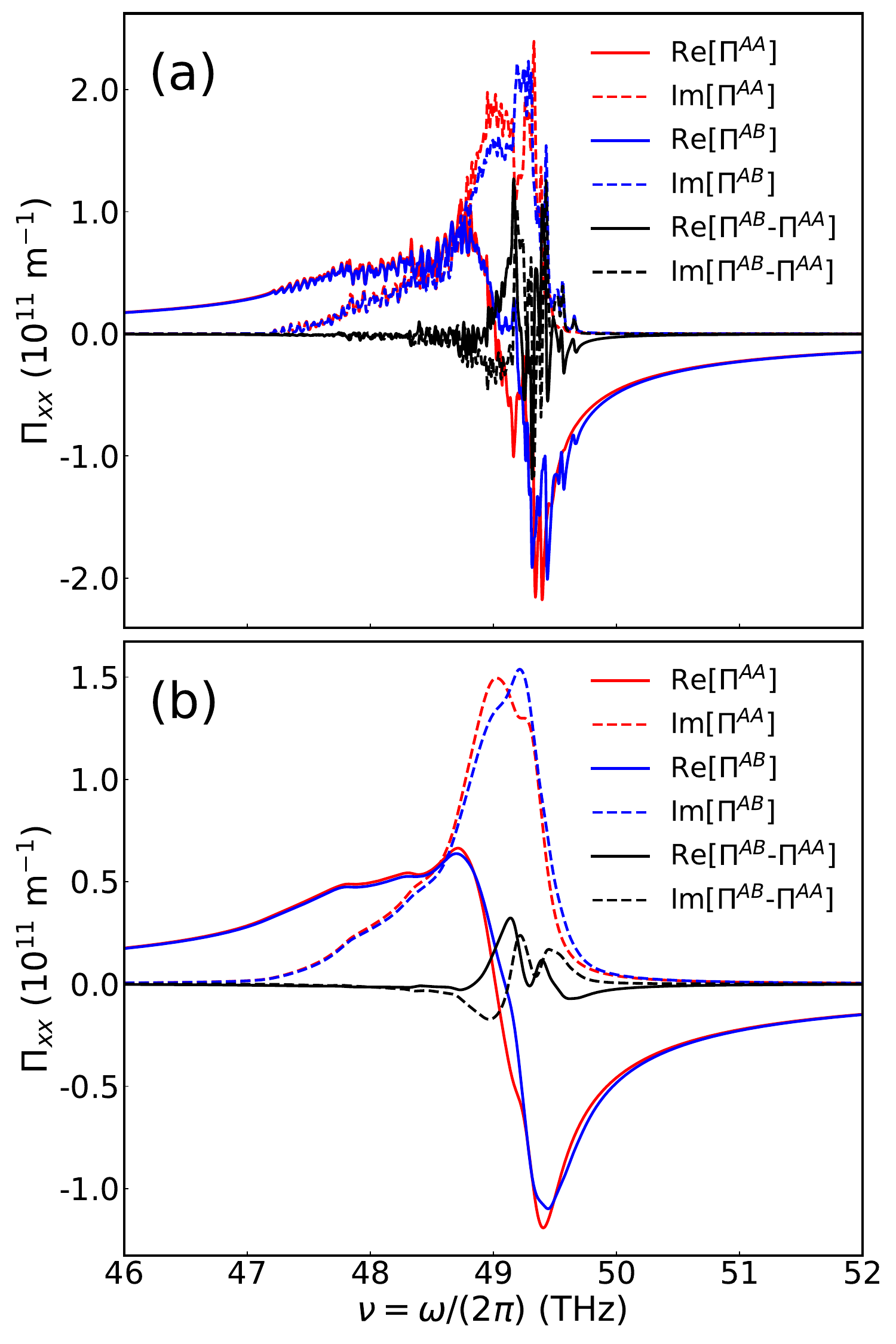}
\caption{\label{fig:fig_local_Pi}The local susceptibility as a function of frequency, calculated using linewidths (a) $\delta/(2\pi)=0.015$ THz ($0.5$ cm$^{-1}$) and (b) $\delta/(2\pi)=0.15$ THz ($5$ cm$^{-1}$). The red and blue curves denote the values at AA and AB points. The solid and dashed lines represent the real and imaginary parts. The black lines indicate their difference.}
\end{figure}

Another key feature of moir\'{e} polar systems is their spatially varying local response. This provides important signatures for detection using scanning near-field optical microscopy (SNOM) \cite{21nc_twisted_hBN_exp,23n_snom_exp_BN_tube}. In SNOM measurements, a tightly focused light field $\bm{E}\sim\delta(\bm{r}-\bm{r}_0)e^{-i\omega t}$ illuminates the sample, and the response at the same position $\bm{r}_0$ is measured. This technique probes the local susceptibility $\bm{\Pi}(\bm{r},\bm{r},\omega)$, which in our formalism can be calculated as ($N_m\Omega_m$ is the sample area)
\begin{align}
\bm{\Pi}(\bm{r},\bm{r},\omega)=\frac{1}{N_m\Omega_m}\sum_{\bar{\bm{q}}\bm{Q}\bm{Q}'}\bm{\Pi}^{\bm{Q}\bm{Q}'}(\bar{\bm{q}},\omega)e^{i(\bm{Q}-\bm{Q}')\cdot\bm{r}}. \label{eq:response_local}
\end{align}
We see that a system can have an inhomogeneous local response, i.e., $\bm{\Pi}(\bm{r},\bm{r},\omega)$ depends explicitly on $\bm{r}$, if and only if $\bm{\Pi}^{\bm{Q}\bm{Q}'}(\bar{\bm{q}},\omega)$ is not diagonal about $\bm{Q}$. This rules out the possibility of observing spatially varying signals in moir\'{e}-free systems such as monolayer hBN. We numerically calculate Eq. (\ref{eq:response_local}) at two different stacking points, AA and AB, using a $7\times 7$ sample mesh of $\bar{\bm{q}}$, 61 truncated $\bm{Q}$ vectors, and two different phonon linewidths $\delta$. The results of $\Pi_{xx}$ in the frequency window $46$-$52$ THz are shown in Fig. \ref{fig:fig_local_Pi} (time reversal and $C_{3z}$ symmetries require $\Pi$ to be proportional to the identity matrix, as shown in SI Section 5.3). In Fig. \ref{fig:fig_local_Pi}(a), using a tiny $\delta$ leads to the sawtooth pattern of $\Pi_{xx}$. Each peak corresponds to a specific moir\'{e} mode. These sharp features are smeared when a larger, more realistic $\delta$ is used, as shown in Fig. \ref{fig:fig_local_Pi}(b). The signal difference between the AA and AB points becomes pronounced in a narrower window ($48.5$-$50$ THz), where moir\'{e} PhPs are active [Fig. \ref{fig:fig_PhP_dispersion}(a)]. Outside this range, the moir\'{e} potential has little effect, and the difference is negligible. Notably, this signal difference persists and remains sizable even under realistic line broadening $\delta$, which is a key characteristic of moir\'{e} polaritons. These numerical results agree qualitatively with previous SNOM experiments \cite{21nc_twisted_hBN_exp}. The spatial variation of near-field response remains robust against linewidth broadening, ensuring reliable experimental detections.

We also calculate the PhP spectrum of $3.89^\circ$ twisted bilayer MoTe$_2$ (SI Section 3.4), which has aroused great interest recently \cite{23n_FQAH_cai,24n_FQAH_kang}. Compared with hBN, the gaps between mini-branches in MoTe$_2$ are smaller, and its critical frequency $\omega_0/(2\pi)\approx 7.2$ THz is also lower, due to the heavier atomic mass. However, some basic properties are qualitatively the same. For example, the spectrum also consists of a linearly dispersive dominant branch and some flat mini-branches, and the intensities become weaker as $\omega$ deviates from $\omega_0$ to lower frequencies.

{\it The continuum model}~~
Finally, we introduce a continuum model that could reproduce the same physics. This model generalizes Huang's continuum Eqs. (\ref{eq:EoM}) and (\ref{eq:P_WE}), and is more computationally efficient than the lattice model since it contains only a few parameters (SI Section 4.2, 4.3). In this model, the vibration field consists of layer- and (commensurate wavevector) $\bm{Q}$-resolved terms: $\bm{W}=\sum_{\bm{Q}l}\bm{W}_{\bm{Q}l}e^{i\bm{Q}\cdot\bm{r}}$. Each $\bm{W}_{\bm{Q}l}$ has a unique resonance frequency $\omega_{\bm{Q}l}$. These components couple to each other and to the electric field: $\ddot{\bm{W}}_{\bm{Q}l}=-\sum_{\bm{Q}'l'}\mathcal{D}_{\bm{Q}l,\bm{Q}'l'}\bm{W}_{\bm{Q}'l'}+\gamma\bm{E}_{\bm{Q}}$, and the polarization is $\bm{P}=\sum_{\bm{Q}l}\gamma\bm{W}_{\bm{Q}l}e^{i\bm{Q}\cdot\bm{r}}$, where $\gamma$ is the same as $\gamma_{12}$ in Eq. (\ref{eq:EoM}). The matrix $\mathcal{D}$ takes nonzero elements only for $|\bm{Q}-\bm{Q}'|\leq \frac{4\pi}{\sqrt{3}L_\theta}$ \cite{11pnas_BM_model}. The diagonal terms $\mathcal{D}_{\bm{Q}l,\bm{Q}l}$ are $\omega_{\bm{Q}l}^2$, and the off-diagonal terms $\mathcal{D}_{\bm{Q}l,\bm{Q}'l'}$ hybridize different components $\bm{W}_{\bm{Q}l}$. To understand why it works, we note that the continuum model essentially describes a system of coupled harmonic oscillators driven by an external field (SI Section 4.1). The elastic coupling (moir\'{e} potential) turns the single-pole susceptibility Eq. (\ref{eq:P_WE}) into the multi-pole one Eq. (\ref{eq:Pi_result}) \cite{21n_exciton_polariton_exp}. This means the long-wavelength optical components are scattered and redistributed among the phonon branches that are backfolded to the mBZ center. In general, the model parameters depend on lattice relaxations in larger supercells, which will be systematically studied in the future.

{\it Summary and outlooks}~~
Following the spirit of Huang's theory, we have derived a set of macroscopic equations to understand 2D PhPs. For moir\'{e} systems, the eigen equation couples different momentum together, resulting in multiple branches of inhomogeneous PhP modes with moir\'{e} patterns. The theoretical proposal has been numerically verified using the lattice model. Many PhP bands are obtained, each carrying a unique EM wave that differs in polarization and localization. The inhomogeneous multi-branch physics can be understood by generalizing Huang's theory to that of coupled harmonic oscillators. In this study, we have calculated only for two specific moir\'{e} systems with relatively small supercells. There remains plenty of room to explore the dependence of optical properties on material parameters. For example, samples with supercells comparable to achievable light wavelengths are more promising for experiments \cite{21nc_twisted_hBN_exp}. The properties of moir\'{e} PhPs could be engineered via the twisting angle, which would conceivably balance the moir\'{e} potential strength against the separation of folded phonon bands—a systematic study of this dependence is an important direction for future studies. The spatial localization of EM waves and the tunability of their wavelength and frequency represent fascinating features of 2D optics. If such modes can be excited efficiently, they could provide flexible driven potentials that differ completely from traditional light fields \cite{21prl_spacetime_crystal,24a_tBN_exp2}. We defer these explorations to future studies.

{\it Supporting information}~~
The supporting information is available online, containing: (I) The derivation of the macroscopic 2D PhP theory; (II) The lattice model of 2D PhP; (III) The lattice model of moir\'{e} PhP; (IV) The macroscopic theory of moir\'{e} PhP; (V) More details about the moir\'{e} response function; (VI) The interatomic force constants used in MD simulations.

\begingroup
\renewcommand{\addcontentsline}[3]{}
\begin{acknowledgments}
We thank X.-D. Guo, W.-Q. Miao, and T.-Y. Qiao for helpful discussions. X. Dai is supported by the New Cornerstone Foundation and a fellowship and a CRF award from the Research Grants Council of the Hong Kong Special Administrative Region, China (Projects No. HKUST SRFS2324-6S01 and No. C7037-22GF). D. Pan is supported by National Natural Science Foundation of China through the Excellent Young Scientists Fund (22022310).
\end{acknowledgments}
\endgroup

\bibliography{reference}

\begin{figure*}
\includegraphics[width=0.43\textwidth]{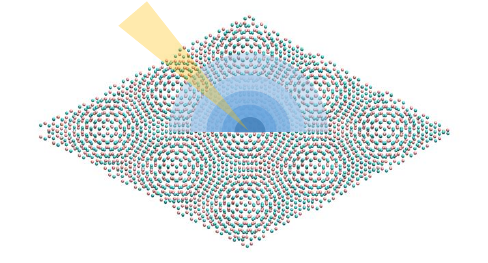}
\caption{\label{TOC Graphic} TOC Graphic.}
\end{figure*}
 
\clearpage 
\newpage 
\onecolumngrid
\begin{center}
\textbf{Supporting Information}\\ 
\end{center}

\renewcommand{\thefigure}{S\arabic{figure}}

\renewcommand{\thetable}{S\arabic{table}}

\renewcommand{\thesection}{S\arabic{section}}

\renewcommand{\theequation}{S\arabic{equation}}

\tableofcontents

\clearpage

\section{Macroscopic theory of 2D PhP}
\subsection{Huang's equation in 2D \label{sec:2D_Huang_app}}
Consider a 2D polar sheet placed at $z=0$ of a dielectric with permittivity $\varepsilon_0\varepsilon$ (in this document we generalize the vacuum case) and permeability $\mu=1$ [Fig. 1(a) of the main text]. We import a continuum vibrational field $\bm{W}$ defined in the plane. For diatomic ionic crystals like hBN it is connected to the ion displacements through $\bm{W}\propto\bm{u}_+-\bm{u}_-$. Under an electric field $\bm{E}$, $\bm{W}$ satisfies the equation of motion of a driven oscillator (neglect dissipation)
\begin{align}
\ddot{\bm{W}}=-\omega_0^2\bm{W}+\gamma_{12}\bm{E}_t, \label{eq:EoM_app}
\end{align}
where $\omega_0$ is the optical phonon frequency resulting from elastic forces, $\bm{E}_t$ is the in-plane part of $\bm{E}$ at $z=0$. The surface polarization density (in-plane dipole moment per unit area) is denoted by $\bm{P}$. In 2D it originates mainly from the relative ionic displacement (rigid-ion approximation)
\begin{align}
\bm{P}=\gamma_{21}\bm{W}. \label{eq:P_WE_app}
\end{align}
The parameters $\gamma_{12}=\gamma_{21}$ due to Onsager relations (or can be inferred from microscopic models). For clarity we set $\gamma_{12}^2=\gamma_{21}^2=\epsilon_0T$. The above two equations are 2D version of Huang's equations. Their microscopic origin is derived in Sec. \ref{sec:microscopic_relation_app}. They should be solved together with Maxwell's equations. These equations govern the form of EM waves in the dielectric $z\neq0$, and, at the interface $z=0$ reduce to boundary conditions (BC):
\begin{subequations}
\begin{align}
&\text{1st}:\quad\bm{E}_t^+-\bm{E}_t^-=\bm{0},\\
&\text{2nd}:\quad \varepsilon_0\varepsilon(E_z^+-E_z^-)=\rho,\\
&\text{3rd}:\quad\bm{B}_t^+-\bm{B}_t^-=\mu_0\bm{J}\times\bm{e}_z,\\
&\text{4th}:\quad B_z^+-B_z^-=0,
\end{align}
\end{subequations}
where $\rho$ and $\bm{J}$ are surface charge and current densities, and `$\pm$' indicates fields just above or below the sheets, e.g., $\bm{E}^\pm=\bm{E}(z=0^{\pm})$. We study solutions of the form 
\begin{align}
\bm{E},\bm{W}\propto e^{i\bm{q}\cdot\bm{r}-i\omega t},
\end{align}
where $\bm{r}=(x,y)$ and $\bm{q}$ are \textit{in-plane} position and momentum, respectively. The in-plane dipole current and charge density
\begin{align}
\bm{J}=\partial\bm{P}/\partial t=-i\omega\bm{P}, \quad \rho=-i\bm{q}\cdot\bm{P}.
\end{align}
Then the 2nd and 3rd BCs become
\begin{align}
&\varepsilon_0\varepsilon(E_z^+-E_z^-)=-i\bm{q}\cdot\bm{P}, \label{eq:2nd_BC_app}\\
&\partial_z\bm{E}_t^+-\partial_z\bm{E}_t^--i\bm{q}(E_z^+-E_z^-)=-\mu_0\omega^2\bm{P},\label{eq:3rd_BC_app}
\end{align}
while the 4th BC coincides with the 1st one. In the dielectric ($z\neq0$), the $\bm{E}$ field obeys the divergence law and the wave equation
\begin{align}
&\nabla\cdot\bm{E}=i\bm{q}\cdot\bm{E}_t+\partial_zE_z=0, \label{eq:divergence_theorem_app} \\
&(\nabla^2-\varepsilon\partial_t^2)\bm{E}=\left(\partial_z^2-q^2+\varepsilon\frac{\omega^2}{c^2}\right)\bm{E}=\bm{0}. \label{eq:wave_equation_app}
\end{align}
Depending on the sign of 
\begin{align}
\lambda^2=q^2-\varepsilon\frac{\omega^2}{c^2},
\end{align}
the solutions can be divided into guided or radiative modes. For $\omega\neq\omega_0$, the susceptibility can be defined
\begin{align}
\bm{P}=\varepsilon_0\Pi(\omega)\bm{E}_t,\quad \Pi(\omega)=\frac{T}{\omega_0^2-\omega^2}. \label{eq:toy_polarization_app}
\end{align}

\subsection{Non-retarded solutions}
Let us first consider the non-retarded (static) limit without \textit{external} field. In this limit, $c$ is taken to infinity, so $\lambda=q$. Furthermore, we only need to consider the 1st and 2nd BCs. At $q=0$, pure oscillation $\bm{W}\neq\bm{0}$ happens and $\bm{E}=\bm{0}$, with the transverse and longitudinal oscillations sharing the same frequency $\omega_0$. For $q>0$, the transverse pure oscillation with frequency $\omega_{\text{TO}}=\omega_0$ is still a solution (TO mode), but the longitudinal mode will move with built-in electric field. For the longitudinal mode, $\bm{E}_t$ has nonzero longitudinal component $E_{\parallel}$. Suppose $E_{\parallel}\propto e^{i\bm{q}\cdot\bm{r}-\lambda|z|-i\omega t}$. In the static limit, Eq. (\ref{eq:divergence_theorem_app}) gives $E_z^{\pm}=\pm iE_{\parallel}$, and the 2nd BC gives $2iE_{\parallel}=-iqP_{\parallel}/(\epsilon_0\varepsilon)$. Combining this with Eq. (\ref{eq:toy_polarization_app}) yields the static longitudinal mode condition: 
\begin{align}
1+\frac{q}{2\varepsilon}\Pi(\omega)=0, \label{eq:static_LO_condition_app}
\end{align}
from which we get the longitudinal dispersion
\begin{align}
\omega_{\text{LO}}=\omega_0\sqrt{1+\frac{qT}{2\varepsilon\omega_0^2}}\approx\omega_0+\frac{qT}{4\varepsilon\omega_0}. \label{eq:LO_dispersion_app}
\end{align}
The static transverse and longitudinal phonon dispersions are shown in Fig. 1(b) of the main text using dashed lines. We see that the above model, although simple, captures both the degeneracy of TO and LO phonons at $\Gamma$ point and the linear LO-TO splitting in the long-wavelength regime, which are typical properties of 2D polar systems \cite{17nl_LOTO_degeneracy,19nl_2D_polariton,24nc_nonanalytic_BN_exp}. We notice that only in the non-retarded limit, the condition (\ref{eq:static_LO_condition_app}) for LO phonon coincides with that for the 2D TM polariton studied before \cite{19nl_2D_polariton,21nl_2D_phonon_oxide}.

\subsection{Guided modes: 2D EM \label{Sec:2DEM_app}}
First let us focus on the regime $\lambda^2>0$. We focus on the case $\omega\neq\omega_0$, where Eq. (\ref{eq:toy_polarization_app}) can be used to eliminate $\bm{W}$ ($\omega=\omega_0$ leads to the trivial solution $\bm{W}=\bm{E}=\bm{0}$). Setting $\lambda=\sqrt{q^2-\varepsilon\omega^2/c^2}>0$, we may assume the following localized $\bm{E}$ field
\begin{align}
  \bm{E}(\bm{r},z,t)= \left\{ \begin{array}{c}
    (\bm{E}_t+E_z^+\bm{e}_z) e^{i\bm{q} \cdot \bm{r} - \lambda z- i\omega t}\quad z>0\\
    (\bm{E}_t+E_z^-\bm{e}_z) e^{i\bm{q} \cdot \bm{r} + \lambda z- i\omega t}\quad z<0
  \end{array} \right.,\label{eq:trial_E_app}
\end{align}
where $\bm{E}_t=E_{\parallel}\bm{e}_{\bm{q}}+E_{\perp}\bm{e}_z\times\bm{e}_{\bm{q}}$. Here, $\bm{e}_{\bm{q}}$ is the unit vector along $\bm{q}$. The divergence theorem gives
\begin{align}
E_z^+=-E_z^-=iqE_{\parallel}/\lambda.
\end{align}
The 2nd BC (\ref{eq:2nd_BC_app}) becomes
\begin{align}
2\frac{iqE_{\parallel}}{\lambda}=-\frac{iqP_{\parallel}}{\epsilon_0\epsilon}\Rightarrow \left[1+\frac{\lambda}{2\varepsilon}\Pi(\omega)\right]E_{\parallel}=0,
\end{align}
The longitudinal part of the 3rd BC (\ref{eq:3rd_BC_app}) is equivalent to the 2nd one, while the transverse part reads
\begin{align}
2\lambda qE_{\perp}=\mu_0\omega^2qP_{\perp}\Rightarrow\left[1-\frac{1}{2\lambda}\frac{\omega^2}{c^2}\Pi(\omega)\right]E_{\perp}=0.
\end{align}
We see that in this isotropic toy model, the solutions for $E_{\parallel}$ and $E_{\perp}$ are decoupled. In realistic models, especially in moir\'{e} materials, they are in general coupled.

Consider first the case $E_{\parallel}\neq0$, $E_{\perp}=0$. This gives a $\bm{B}$ field polarized in the direction perpendicular to $\bm{e}_{\bm{q}}$, 
\begin{align}
\begin{split}
&\bm{E}^{\pm}=\left(E_{\parallel}\bm{e}_{\bm{q}}\pm i\frac{q}{\lambda}E_{\parallel}\bm{e}_z\right)e^{i\bm{q}\cdot\bm{r}-i\omega t-\lambda|z|}, \\
&\bm{B}^{\pm}=\mp i\frac{\omega\varepsilon}{\lambda c^2}E_{\parallel}\bm{e}_z\times\bm{e}_{\bm{q}}e^{i\bm{q}\cdot\bm{r}-i\omega t-\lambda|z|},
\end{split}
\end{align}
which is the transverse magnetic (TM) mode. In this case this eigen equation reduces to the TM mode condition
\begin{align}
\text{TM} : \quad 1+\frac{\lambda}{2\varepsilon}\Pi(\omega)=0. \label{eq:TM_condition_app}
\end{align}
Since $\lambda>0$, it has a solution when $\Pi(\omega)<0$, i.e., when $\omega>\omega_0$. The dispersion reads
\begin{align}
q=\sqrt{\varepsilon\frac{\omega^2}{c^2}+\varepsilon^2\frac{(\omega^2-\omega_0^2)^2}{T^2/4}}.
\end{align}
At $\omega=\omega_0$, $q=q_0=\sqrt{\varepsilon}\omega_0/c$, the group velocity $v_0=\text{d}\omega/\text{d}q|_{q=q_0}=c/\sqrt{\varepsilon}$, i.e., the TM mode is tangential to the light cone. As shown in Fig. 1(b), when $q\gg q_0$, the dispersion becomes linear and asymptotically approaches the static LO mode Eq. (\ref{eq:LO_dispersion_app}).

Then consider the case $E_{\parallel}=0$, $E_{\perp}\neq0$, which corresponds to the transverse electric (TE) mode
\begin{align}
\begin{split}
&\bm{E}^{\pm}= E_{\perp}\bm{e}_z\times\bm{e}_{\bm{q}} e^{i\bm{q}\cdot\bm{r}-i\omega t-\lambda|z|},\\
&\bm{B}^{\pm}=\left(\pm\frac{\lambda}{i\omega}E_{\perp}\bm{e}_{\bm{q}}+\frac{q}{\omega}E_{\perp}\bm{e}_z\right)e^{i\bm{q}\cdot\bm{r}-i\omega t-\lambda|z|}.
\end{split}
\end{align}
The dispersion obeys the TE mode condition
\begin{align}
\text{TE} :\quad 1-\frac{1}{2\lambda}\frac{\omega^2}{c^2}\Pi(\omega)=0, \label{eq:TE_condition_app}
\end{align}
which has solutions when $\Pi(\omega)>0$, i.e., when $\omega<\omega_0$. The dispersion reads
\begin{align}
q=\frac{\omega}{c}\sqrt{\varepsilon+\frac{\omega^2}{c^2}\frac{T^2/4}{(\omega_0^2-\omega^2)^2}}.
\end{align}
When $q<q_0$, the TE mode resembles light: the dispersion closely follows the light cone, hence with a tiny $\lambda$ and weak localization. When $q>q_0$, it turns almost into pure lattice oscillations: the dispersion remains very close to the static TO line $\omega=\omega_0$ and the EM fields are weak and extremely localized at the surface.

The above TM and TE modes are typical 2D EM waves with their energy constrained along the $z$-axis. Such modes, accompanied by polarizable collective modes, exist ubiquitously in 2D materials or at the interfaces of 3D materials. The eigenmode conditions (\ref{eq:TM_condition_app}) and (\ref{eq:TE_condition_app}) are general. For example, when discussing 2D plasmon polaritons one only needs to replace $\Pi(\omega)$ by its plasmon version. It is also true that the sign of $\Pi(\omega)$ governs whether the mode is TM or TE.

\subsection{Radiative solutions: 3D EM \label{Sec:3DEM_app}}
Then we consider the case $\lambda^2<0$, i.e., the left side of the light cone. This means the wave vector along the $z$-axis is real, and we take $\lambda=-i\sqrt{\varepsilon\omega^2/c^2-q^2}=-ik_z$. In other words, EM fields occupy the whole 3D space without decaying. In this case, we can no longer expect unidirectional waves, like those in Eq. (\ref{eq:trial_E_app}), to exist in both $z>0$ and $z<0$ regions. Otherwise, if $\bm{E}^{\pm}=(\bm{E}_t+E_z^{\pm}\bm{e}_z)e^{i\bm{q}\cdot\bm{r}\pm ik_zz-i\omega t}$, following the derivation for $\lambda^2>0$ case, the resulting eigen equations are the same as Eqs. (\ref{eq:TM_condition_app}) and (\ref{eq:TE_condition_app}), which have no solution since $\lambda$ is imaginary. Alternatively, if we take $\bm{E}^{\pm}=(\bm{E}_t+E_z^{\pm}\bm{e}_z)e^{i\bm{q}\cdot\bm{r}+ik_zz-i\omega t}$, we find that $E_z^+=E_z^-$. The 3rd BC gives $\bm{P}=\bm{0}$ and thus $\bm{W}=\bm{E}=\bm{0}$, which is also trivial. Instead, the $\bm{E}$ field must be composed of waves propagating in multiple directions in at least one half-space. A typical solution is the light incidence setup, where the 2D sheet is treated as a scattering potential. In the half-space containing the light source, plane waves propagate in two directions (incident and reflected). For this setup, a non-trivial solution exists for all $(\bm{q},\omega)$ except when $\omega=\omega_0$, corresponding to the continuous spectrum in the dispersion plots. We will solve these in the next subsection, where the light incidence setup is generalized to contain both guided and radiative modes.

\subsection{Revisit 2D PhP as a light reflection and refraction problem \label{subsec:light_incidence_app}}
Both the localized and radiative modes can be understood from the light reflection and refraction viewpoint. This approach is directly connected to experimental techniques for exciting these modes and offers greater numerical convenience for revealing their dispersions. We denote the incident, reflected, and refracted light as $\bm{E}^i$, $\bm{E}^r$, and $\bm{E}^t$. Then we assume the light is incident from $z<0$,
\begin{align}
\begin{split}
&\bm{E}^i=(\bm{E}^i_t+iqE_{\parallel}^i/\lambda\bm{e}_z)e^{i\bm{q}\cdot\bm{r}-\lambda z-i\omega t},\\
&\bm{E}^r=(\bm{E}^r_t-iqE_{\parallel}^r/\lambda\bm{e}_z)e^{i\bm{q}\cdot\bm{r}+\lambda z-i\omega t},\\
&\bm{E}^t=(\bm{E}^t_t+iqE_{\parallel}^t/\lambda\bm{e}_z)e^{i\bm{q}\cdot\bm{r}-\lambda z-i\omega t},
\end{split}
\end{align}
where the in-plane part $\bm{E}_t^l=E_{\parallel}^l\bm{e}_{\bm{q}}+E_{\perp}^l\bm{e}_z\times\bm{e}_{\bm{q}}$, $l=i,r,t$. Here $\lambda$ is allowed to take real or imaginary values,
\begin{align}
\lambda= \left\{ \begin{array}{c}
    -i\sqrt{\varepsilon\frac{\omega^2}{c^2}-q^2},\quad q<\sqrt{\varepsilon}\omega/c\\
    \sqrt{q^2-\varepsilon\frac{\omega^2}{c^2}},\quad\quad q>\sqrt{\varepsilon}\omega/c
  \end{array} \right., \label{eq:lambda_z_app}
\end{align}
depending on whether the EM wave is radiative or guided. Note that for the guided case ($q > \sqrt{\varepsilon}\omega/c$), an incident wave from a prism is also evanescent (attenuated). Such a wave can be generated using the Otto configuration \cite{16ws_book_plasmon}. Then the 1st BC reads
\begin{align}
\bm{E}^t_t=\bm{E}^i_t+\bm{E}^r_t.
\end{align}
The 2nd BC reads
\begin{align}
\bm{e}_{\bm{q}}\cdot\left[\bm{E}_t^t-\bm{E}_t^i+\bm{E}_t^r+\frac{\lambda}{\varepsilon}\Pi(\omega)\bm{E}_t^t\right]=0. \label{eq:longitudinal_monolayer_app}
\end{align}
The longitudinal part of the 3rd BC is still equivalent to the above one, and the transverse part reads
\begin{align}
\bm{e}_{\bm{q}}\times\left[\bm{E}_t^t-\bm{E}_t^i+\bm{E}_t^r-\frac{1}{\lambda}\frac{\omega^2}{c^2}\Pi(\omega)\bm{E}_t^t\right]=0. \label{eq:transverse_monolayer_app}
\end{align}
From the above equations, we can express $\bm{E}^t$ and $\bm{E}^r$ using the incident light $\bm{E}^i$,
\begin{align}
\bm{E}_t^t=T(\bm{q},\omega)\bm{E}_t^i,\quad\bm{E}_t^r=R(\bm{q},\omega)\bm{E}_t^i,
\end{align}
and the transmission and reflection matrices are (written in the basis of $E_{\parallel}^{l}$, $E_{\perp}^l$, where $I$ is the identity matrix)
\begin{subequations}
\begin{align}
&T(\bm{q},\omega)=\left(\begin{matrix} 1+\frac{\lambda}{2\varepsilon}\Pi(\omega)& \\ & 1-\frac{1}{2\lambda}\frac{\omega^2}{c^2}\Pi(\omega)\end{matrix}\right)^{-1},\\
&R(\bm{q},\omega)=T(\bm{q},\omega)-I.
\end{align}
\end{subequations}

When $(\bm{q},\omega)$ lies in the radiative regime, the above equations describe the usual transmission and reflection problem discussed in Sec. \ref{Sec:3DEM_app}. When $(\bm{q},\omega)$ lies in the guided regime, the matrices $T$ and $R$ have poles corresponding exactly to the polariton dispersion discussed in Sec. \ref{Sec:2DEM_app}. Poles indicate that $E^t$ and $E^r$ can be induced with an infinitesimal incidence $E^i$, indicating the excitation of intrinsic modes. We can visualize both the continuous spectrum and the discrete dispersion by plotting the transmission spectrum 
\begin{align}
\mathcal{L}(\bm{q},\omega)=-\text{Im}[\text{det}[T(\bm{q},\omega+i\delta/2)]],
\end{align}
where the phonon linewidth $\delta$ is tiny and positive.

In this section we discuss only the moir\'{e}-less case. In moir\'{e} materials or heterostructures, multiple-pole response should be incorporated into Eqs. (\ref{eq:EoM_app}) and (\ref{eq:P_WE_app}) to correctly describe the long-wavelength behavior. We leave such a generalization for Sec. \ref{sec:append_coupled_oscillator_app}.

\section{Lattice model of 2D PhP in simple polar systems}
\subsection{Lattice dynamics of 2D polar systems}
In this section we build the lattice theory for 2D polar systems. We focus on the moir\'{e}-less case here, but will generalize it to include moir\'{e} effects in the next section. In such a model, the short-ranged force among ions is described by the force constant $\Phi_{\alpha,\beta}$ introduced in Sec. \ref{sec_IFC_app}, while the long-ranged Coulomb force (specifically, dipole-dipole interaction) is incorporated into the macroscopic electric fields. In the long-wavelength limit, it should reduce to the macroscopic Huang's theory introduced in Sec. \ref{sec:2D_Huang_app}. 

Suppose we have a lattice where the ions' equilibrium positions are $\bm{r}_{i\alpha}=\bm{R}_i+\bm{\tau}_{\alpha}$, where $\bm{R}_i$ denotes the $i$-th unit cell, $\bm{\tau}_{\alpha}$ denotes the relative position of sublattice $\alpha$. With the long-wavelength electric field $\bm{E}(\bm{r})=\bm{E}_{\bm{q}}(\omega)e^{i\bm{q}\cdot\bm{r}-i\omega t}$ ($\bm{q}\approx\bm{0}$), the equation of motion for lattice displacement $\bm{u}(\bm{r}_{i\alpha})$ is (we focus on in-plane dynamics in this study, so $\mu,\nu$ take values in $x,y$ only)
\begin{align}
M_{\alpha}\ddot{u}_{\mu}(\bm{r}_{i\alpha})+\sum_{j\beta\nu}\Phi_{\alpha\mu,\beta\nu}(\bm{r}_{i\alpha}-\bm{r}_{j\beta})u_{\nu}(\bm{r}_{j\beta})-Z_{\alpha}eE_{\bm{q},\mu}(\omega)e^{i\bm{q}\cdot\bm{r}_{i\alpha}-i\omega t}=0, \label{eq:EoM_moireless_app}
\end{align}
where $Z_{\alpha}$ and $M_{\alpha}$ are effective (dimensionless) charge and mass (which are assumed isotropic) of sublattice $\alpha$, $e$ is the elementary charge, and $\Phi_{\alpha,\beta}$ is the force constant matrix for nearby ions. In the absence of $\bm{E}$, the above equation reduces to the harmonic equation of motion in the usual phonon problem. The polarization density is defined as \cite{96oxford_book,18a_linear_response_Pi}
\begin{align}
\bm{P}(\bm{r})=\sum_{i\alpha}Z_{\alpha}e\bm{u}(\bm{r}_{i\alpha})\delta(\bm{r}-\bm{r}_{i\alpha}), \label{eq:P_simple_app}
\end{align}
where $\delta(\bm{r}-\bm{r}_0)$ is the 2D Dirac delta function. The above two equations are the lattice version of Eqs. (\ref{eq:EoM_app}), (\ref{eq:P_WE_app}). The linearity of the oscillators allows for an analytical solution of $\bm{u}(\bm{r}_{i\alpha})$ in the presence of the $\bm{E}$ field. Suppose $\bm{u}$ can be expanded as
\begin{align}
\bm{u}(\bm{r}_{i\alpha})=\sum_{a}\frac{1}{\sqrt{M_{\alpha}}}e^{i\bm{q}\cdot\bm{r}_{i\alpha}-i\omega t}\bm{e}_{\alpha,a}(\bm{q})B_a(\bm{q},\omega), \label{eq:trial_u_moireless_app}
\end{align}
where $\bm{e}_a(\bm{q})$ is the displacement vector of the $a$-th eigenmode with frequency $\omega_{\bm{q}a}$, satisfying
\begin{align}
\sum_{\beta\nu}D_{\alpha\mu,\beta\nu}(\bm{q})e_{\beta\nu,a}(\bm{q})=\omega_{\bm{q}a}^2e_{\alpha\mu,a}(\bm{q}),\label{eq:moireless_EoM_app}
\end{align}
and the orthogonality
\begin{align}
\sum_{\alpha\mu}e_{\alpha\mu,a}^*(\bm{q})e_{\alpha\mu,a'}(\bm{q})=\delta_{aa'},\quad
\sum_{a}e^*_{\alpha\mu,a}(\bm{q})e_{\beta\nu,a}(\bm{q})=\delta_{\alpha\beta}\delta_{\mu\nu}, \label{eq:moireless_orthogonality_app}
\end{align}
where the $k$-space dynamical matrix reads
\begin{align}
D_{\alpha,\beta}(\bm{q})=\sum_j\frac{\Phi_{\alpha,\beta}(\bm{r}_{i\alpha}-\bm{r}_{j\beta})}{\sqrt{M_{\alpha}M_{\beta}}}e^{i\bm{q}\cdot(\bm{r}_{j\beta}-\bm{r}_{i\alpha})}. 
\end{align}
The goal is to solve $B_a(\bm{q},\omega)$. Plugging Eq. (\ref{eq:trial_u_moireless_app}) into Eq. (\ref{eq:EoM_moireless_app}), we get
\begin{align}
-\omega^2\sum_{a}e_{\alpha\mu,a}(\bm{q})B_{a}(\bm{q},\omega)+\sum_{a\beta\nu}D_{\alpha\mu,\beta\nu}(\bm{q})e_{\beta\nu,a}(\bm{q})B_a(\bm{q},\omega)-\frac{Z_{\alpha}e}{\sqrt{M_{\alpha}}}E_{\bm{q},\mu}=0. \label{eq:EoM_complex_app}
\end{align}
Then using Eqs. (\ref{eq:moireless_EoM_app}), (\ref{eq:moireless_orthogonality_app}), we get [recall that $\bm{E}_t$ is the in-plane part of $\bm{E}(z=0)$]
\begin{align}
(\omega_{\bm{q}a}^2-\omega^2)B_a(\bm{q},\omega)=e\bm{S}^*_{a}(\bm{q})\cdot\bm{E}_{\bm{q},t}, \label{eq:a_expression_moireless_app}
\end{align}
where we have defined the $S$ matrix
\begin{align}
\bm{S}_{a}(\bm{q})=\sum_{\alpha}\frac{Z_{\alpha}\bm{e}_{\alpha,a}(\bm{q})}{\sqrt{M_{\alpha}}}.\label{eq:Smat_moireless_app}
\end{align}
When $\omega\neq\omega_0$ (which is guaranteed in numerics by introducing a tiny linewidth $\omega+i\delta/2$), we can solve for $B_a(\bm{q},\omega)$ and plug it into Eq. (\ref{eq:trial_u_moireless_app}) to obtain $\bm{u}$.

To apply macroscopic Maxwell's equations, we need to derive the continuous field $\bm{P}(\bm{r})$ from the lattice version in Eq. (\ref{eq:P_simple_app}). This process is done by expanding Eq. (\ref{eq:P_simple_app}) in a Fourier series $\bm{P}_{\bm{q}+\bm{b}}$ ($\bm{b}$ spans the reciprocal lattice) \cite{96oxford_book} and retaining only the leading term with $\bm{b}=\bm{0}$. Terms with $\bm{b}\neq\bm{0}$ are redundant because they detail the information within atomic unit cell and are thus not responsible for long-wavelength physics. So
\begin{align}
\bm{P}(\bm{r})=\sum_{\bm{b}}\bm{P}_{\bm{q}+\bm{b}}(\omega)e^{i(\bm{q}+\bm{b})\cdot\bm{r}-i\omega t}\rightarrow \bm{P}(\bm{r})=\bm{P}_{\bm{q}}(\omega)e^{i\bm{q}\cdot\bm{r}-i\omega t}. \label{eq:P_continnum_moireless_app}
\end{align}
Then $\bm{P}_{\bm{q}}(\omega)$ can be calculated as
\begin{align}
\begin{split}
\bm{P}_{\bm{q}}(\omega)=&\frac{1}{N_{\text{tot}}\Omega_0}\int d^2\bm{r}\bm{P}(\bm{r})e^{-i\bm{q}\cdot\bm{r}+i\omega t}=\frac{1}{N_{\text{tot}}\Omega_0}\sum_{i\alpha}Z_{\alpha}e\bm{u}(\bm{r}_{i\alpha})e^{-i\bm{q}\cdot\bm{r}_{i\alpha}+i\omega t}=\frac{e}{\Omega_0}\sum_{a}\bm{S}_{a}(\bm{q})B_a(\bm{q},\omega), \label{eq:Pi_q_cal_moireless_app}
\end{split}
\end{align}
where $\Omega_0$ is the area of the atomic unit cell and $N_{\text{tot}}$ is the total number of unit cells. Using Eq. (\ref{eq:a_expression_moireless_app}), we obtain
\begin{align}
P_{\bm{q},\mu}(\omega)=\varepsilon_0\sum_{\nu}\Pi_{\mu\nu}(\bm{q},\omega)E_{\bm{q},\nu}(\omega),
\end{align}
where the in-plane susceptibility is a 2 by 2 matrix,
\begin{align}
\Pi_{\mu\nu}(\bm{q},\omega)=\frac{e^2}{\varepsilon_0\Omega_0}\sum_a\frac{[\bm{S}_a(\bm{q})]_{\mu}[\bm{S}^*_{a}(\bm{q})]_{\nu}}{\omega^2_{\bm{q}a}-\omega^2}. \label{eq:Pi_result_moireless_app} 
\end{align}
This result is also valid in the quantum case. By treating phonons as bosons, we will rederive this result for moir\'{e} systems using quantum linear response theory in Sec. \ref{append:Response_quantum_app}. All the derivations of PhP conditions follow exactly the same procedure as in Secs. \ref{Sec:2DEM_app}, \ref{Sec:3DEM_app}, \ref{subsec:light_incidence_app}, since the macroscopic response function has already been obtained.

\subsection{Relation to the macroscopic theory \label{sec:microscopic_relation_app}}
Now it is time to relate the lattice model to the macroscopic model introduced in Sec. \ref{sec:2D_Huang_app}. We still focus on the moir\'{e}-less case. In the long-wavelength limit $|\bm{q}|\sim\omega_0/c$, the response function is almost isotropic and dispersionless, i.e., 
\begin{align}
\Pi_{\mu\nu}(\bm{q},\omega)\approx\delta_{\mu\nu}\Pi(\omega).
\end{align}
The nonzero contribution comes from the LO and TO modes (for acoustic modes, the polarization from opposite ions cancels due to in-phase oscillation), which gives the isotropic response
\begin{align}
\Pi(\omega)=\frac{e^2}{\varepsilon_0\Omega_0}\lim_{\bm{q}\rightarrow\bm{0}}\frac{|\bm{e}_{\bm{q}}\cdot\bm{S}_{\text{LO}}(\bm{q})|^2}{\omega_{\bm{q},\text{LO}}^2-\omega^2}=\frac{e^2}{\varepsilon_0\Omega_0}\lim_{\bm{q}\rightarrow\bm{0}}\frac{|(\bm{e}_z\times\bm{e}_{\bm{q}})\cdot\bm{S}_{\text{TO}}(\bm{q})|^2}{\omega_{\bm{q},\text{TO}}^2-\omega^2}.
\end{align}
Compared with Eq. (\ref{eq:toy_polarization_app}), we recognize
\begin{subequations}
\begin{align}
&\omega_0=\omega_{\bm{0},\text{LO}}=\omega_{\bm{0},\text{TO}},\\
&T=\frac{e^2}{\varepsilon_0\Omega_0}\lim_{\bm{q}\rightarrow\bm{0}}\left[\bm{e}_{\bm{q}}\cdot\bm{S}_{\text{LO}}(\bm{q})\right]^2=\frac{e^2}{\varepsilon_0\Omega_0}\lim_{\bm{q}\rightarrow\bm{0}}\left[(\bm{e}_z\times\bm{e}_{\bm{q}})\cdot\bm{S}_{\text{TO}}(\bm{q})\right]^2. \label{eq:T_lattice_to_macro_app}
\end{align}
\end{subequations}
Substituting the above expressions into the static LO dispersion Eq. (\ref{eq:LO_dispersion_app}), we obtain the LO-TO splitting (the so-called non-analytical correction) in exact agreement with that in Ref. \cite{17nl_LOTO_degeneracy}:
\begin{align}
\omega^2(\bm{q})-\omega_0^2=V(\bm{q})\frac{q^2}{\Omega_0}|\bm{e}_{\bm{q}}\cdot\bm{S}_{\text{LO}}(\bm{q})|^2_{\bm{q}\rightarrow\bm{0}},
\end{align}
with the 2D Coulomb interaction $V(\bm{q})=e^2/(2\varepsilon_0\varepsilon q)$. The screening term is missing here because we have used the zero-thickness approximation from the beginning \cite{16prb_Frohlich_2D_reff}. From now on, for simplicity we focus on the monolayer hBN system. At $\bm{q}\approx\bm{0}$, for hBN the LO and TO phonons' eigenmode displacement vectors can be obtained using the invariance of the center of mass: $M_N\bm{u}_N+M_B\bm{u}_B\propto \sqrt{M_N}\bm{e}_N+\sqrt{M_B}\bm{e}_B=0$, so
\begin{align}
\begin{split}
&\bm{e}_{\text{LO}}(\bm{q})=[\bm{e}_{N,\text{LO}}(\bm{q}),\bm{e}_{B,\text{LO}}(\bm{q})]\approx\left(\sqrt{\frac{M_B}{M_N+M_B}}\bm{e}_{\bm{q}},-\sqrt{\frac{M_N}{M_N+M_B}}\bm{e}_{\bm{q}}\right)^T,\\
&\bm{e}_{\text{TO}}(\bm{q})=[\bm{e}_{N,\text{TO}}(\bm{q}),\bm{e}_{B,\text{TO}}(\bm{q})]\approx\left(\sqrt{\frac{M_B}{M_N+M_B}}\bm{e}_z\times\bm{e}_{\bm{q}},-\sqrt{\frac{M_N}{M_N+M_B}}\bm{e}_z\times\bm{e}_{\bm{q}}\right)^T. \label{eq:LOTO_express_app}
\end{split}
\end{align}
At $\bm{q}=\bm{0}$, $\bm{e}_{\bm{q}}=\bm{q}/|\bm{q}|$ is ill-defined, but we can fix $\bm{e}_{\bm{0}}=\bm{e}_x$, $\bm{e}_z\times\bm{e}_{\bm{0}}=\bm{e}_y$, i.e.,
\begin{align}
\bm{e}_{\text{LO}}(\bm{0})=\left(\sqrt{\frac{M_B}{M_N+M_B}}\bm{e}_{x},-\sqrt{\frac{M_N}{M_N+M_B}}\bm{e}_{x}\right)^T,
\bm{e}_{\text{TO}}(\bm{0})=\left(\sqrt{\frac{M_B}{M_N+M_B}}\bm{e}_{y},-\sqrt{\frac{M_N}{M_N+M_B}}\bm{e}_{y}\right)^T. \label{eq:LOTO_at_0_express_app}
\end{align}
For monolayer hBN, the $T$ parameter can then be analytically derived as
\begin{align}
&\lim_{\bm{q}\rightarrow\bm{0}}\bm{e}_{\bm{q}}\cdot\bm{S}_{\text{LO}}(\bm{q})=\lim_{\bm{q}\rightarrow\bm{0}}(\bm{e}_z\times\bm{e}_{\bm{q}})\cdot\bm{S}_{\text{TO}}(\bm{q})=\frac{Z_N}{\sqrt{M_N}}\sqrt{\frac{M_B}{M_N+M_B}}-\frac{Z_B}{\sqrt{M_B}}\sqrt{\frac{M_N}{M_N+M_B}},\\
&T=\frac{e^2}{\varepsilon_0\Omega_0}\frac{M_N M_B}{M_N+M_B}\left(\frac{Z_N}{M_N}-\frac{Z_B}{M_B}\right)^2=3.842\times 10^{19}\text{ m/s}^2.
\end{align}
Using $\omega_0\approx 2\pi\times 49.4463$ THz, we get $T/(2\omega_0 c)\approx 2.06\times 10^{-4}$, which is used to plot Fig. 1(b) in the main text. For the bilayer hBN without moir\'{e} potential (i.e., with no twist), $T$ is doubled. Further, the macroscopic $\bm{W}$ field is related to the lattice dynamics through
\begin{align}
\bm{W}(\bm{r},t)=\frac{1}{\sqrt{\Omega_0}}\sqrt{\frac{M_NM_B}{M_N+M_B}}[\bm{u}_{N}(\bm{r},t)-\bm{u}_{B}(\bm{r},t)]\propto\frac{1}{\sqrt{\Omega_0}}\sqrt{\frac{M_NM_B}{M_N+M_B}}\left[\frac{\bm{e}_{N}(\bm{q})}{\sqrt{M_N}}-\frac{\bm{e}_{B}(\bm{q})}{\sqrt{M_B}}\right]e^{i\bm{q}\cdot\bm{r}-i\omega t}, \label{eq:W_microscopic_app}
\end{align}
where $\bm{u}_{\alpha}(\bm{r},t)$ is the continuum version [similar to Eq. (\ref{eq:P_continnum_moireless_app})] of the displacement field for sublattice $\alpha$. 

We now derive the continuum Huang's equations Eqs. (\ref{eq:EoM_app}), (\ref{eq:P_WE_app}). The key point is that, under the electric field, the optically active displacement field can be expanded using the field-free iLO/iTO modes [Eq. (\ref{eq:trial_u_moireless_app})], i.e., 
\begin{align}
\bm{u}(\bm{r}_{i\alpha})=\frac{e^{i\bm{q}\cdot\bm{r}_{i\alpha}}}{\sqrt{M_{\alpha}}}(\bm{e}_{\alpha,\text{LO}}(\bm{q}),\bm{e}_{\alpha,\text{TO}}(\bm{q}))\left(\begin{matrix}A_{\text{LO}}\\A_{\text{TO}}\end{matrix}\right).
\end{align}
Notice that we have absorbed the time dependence into $A_{\text{LO/TO}}$. Then from Eq. (\ref{eq:EoM_complex_app}) we know
\begin{align}
\begin{split}
\frac{Z_{\alpha}e}{M_{\alpha}}\bm{E}_{\bm{q}}e^{i\bm{q}\cdot\bm{r}_{i\alpha}-i\omega t}
=&\sum_{a}\frac{e^{i\bm{q}\cdot\bm{r}_{i\alpha}}}{\sqrt{M_{\alpha}}}\bm{e}_{\alpha,a}(\bm{q})\ddot{A_{a}}+\sum_{a}\omega_{\bm{q}a}^2\frac{e^{i\bm{q}\cdot\bm{r}_{i\alpha}}}{\sqrt{M_{\alpha}}}\bm{e}_{\alpha,a}(\bm{q})A_a\\
\approx&\sum_{a}\frac{e^{i\bm{q}\cdot\bm{r}_{i\alpha}}}{\sqrt{M_{\alpha}}}\bm{e}_{\alpha,a}(\bm{q})\ddot{A_{a}}+\omega_0^2\sum_{a}\frac{e^{i\bm{q}\cdot\bm{r}_{i\alpha}}}{\sqrt{M_{\alpha}}}\bm{e}_{\alpha,a}(\bm{q})A_a\\
=&\ddot{\bm{u}}(\bm{r}_{i\alpha})+\omega_0^2\bm{u}(\bm{r}_{i\alpha}),
\end{split}
\end{align}
where we have adopted the Einstein approximation: $\omega_{\bm{q}a}=\omega_0$ ($\omega_0$ is the degenerate eigen frequency of iLO/iTO modes at $\bm{q}=\bm{0}$) in the second line. The continuum version of the above equation of motion is simply $\ddot{\bm{u}}_{\alpha}(\bm{r},t)=-\omega_0^2\bm{u}_{\alpha}(\bm{r},t)+(Z_{\alpha}e/M_{\alpha})\bm{E}_t(\bm{r},t)$. Then using the relation Eq. (\ref{eq:W_microscopic_app}) we get Eq. (\ref{eq:EoM_app})
\begin{align}
\ddot{\bm{W}}(\bm{r},t)=-\omega_0^2\bm{W}(\bm{r},t)+\frac{e}{\sqrt{\Omega_0}}\sqrt{\frac{M_NM_B}{M_N+M_B}}\left(\frac{Z_N}{M_N}-\frac{Z_B}{M_B}\right)\bm{E}_t(\bm{r},t).
\end{align}
The relationship Eq. (\ref{eq:P_WE_app}) is easily obtained through the continuum version of Eq. (\ref{eq:P_simple_app})
\begin{align}
\bm{P}(\bm{r},t)=\frac{1}{\Omega_0}\sum_{\alpha}Z_{\alpha}e\bm{u}_{\alpha}(\bm{r},t)=\frac{Z_Ne}{\Omega_0}(\bm{u}_N-\bm{u}_B)=\frac{Z_Ne}{\sqrt{\Omega_0}}\sqrt{\frac{1}{M_N}+\frac{1}{M_B}}\bm{W}(\bm{r},t).
\end{align}
So we see the microscopic expressions for $\gamma_{12}$, $\gamma_{21}$ appearing in Eqs. (\ref{eq:EoM_app}), (\ref{eq:P_WE_app}) are
\begin{align}
\gamma_{12}=\frac{e}{\sqrt{\Omega_0}}\sqrt{\frac{M_NM_B}{M_N+M_B}}\left(\frac{Z_N}{M_N}-\frac{Z_B}{M_B}\right)=\frac{Z_Ne}{\sqrt{\Omega_0}}\sqrt{\frac{1}{M_N}+\frac{1}{M_B}}=\gamma_{21}, \label{eq:gamma_microscopic_app}
\end{align}
and $\gamma_{12}^2=\gamma_{21}^2=\varepsilon_0T$ is recovered. We notice that in deriving the continuum theory, the $\bm{q}$-dependence in the phonon level has been abandoned, which is an excellent approximation since the optical region $q\sim q_0=\omega_0/c$ is four to five orders smaller than the Brillouin zone size $1/a_0$. In other words, the $\bm{q}$-dispersion of PhP comes almost completely from light. We also note that for the moir\'{e} case, when Eq. (\ref{eq:scale_app}) is satisfied, we can still abandon the $\bar{\bm{q}}$-dependence. However, it is necessary to retain the $\bm{Q}$-dependence of $\omega_{\bar{\bm{q}}+\bm{Q},b}$ (and dynamical matrix): $\omega_{\bar{\bm{q}}+\bm{Q},b}\approx \omega_{\bm{Q},b}$. It is such splitting of $\omega_{\bm{Q},b}$ that gives rise to the multiple branches of moir\'{e} PhP.

\subsection{Force constant of monolayer hBN}
Here we detail the lattice model used for monolayer hBN. The short-ranged elastic force constants (FC) $\Phi_{\alpha,\beta}(\bm{r}_{i\alpha}-\bm{r}_{j\beta})$ are used to generate the bare phonon dispersion and eigenmode displacement vectors. Moreover, this monolayer model provides the moir\'{e}-less basis in the continuum model we will derive in Secs. \ref{sec:continuum_phonon_app} and \ref{sec:continuum_polariton_app}.

The monolayer hBN is a hexagonal lattice with lattice constant $a_0=2.504$ \AA. The Bravais lattice vectors are $\bm{a}_1=a_0(1/2,\sqrt{3}/2)$, $\bm{a}_2=a_0(-1/2,\sqrt{3}/2)$. The nitride ($N$) and boron ($B$) atoms are located at $\bm{\tau}_N=-\bm{\tau}_B=(\bm{a}_1+\bm{a}_2)/3=(a_0/\sqrt{3})\bm{e}_y$ near the origin. The lattice has the $C_{3z}$, $C_{2y}$, and $M_z$ symmetries. Notice that $M_z$ results in the decoupling of in-plane and out-of-plane phonons of hBN monolayer, so we can consider only the in-plane parts. For simplicity we retain only the onsite, nearest-neighboring (n.n.), and next-nearest neighboring (n.n.n.) FCs. Each $N$ ($B$) atom has 3 n.n. $B$ ($N$) atoms, denoted by 3 relative vectors $(\bm{r}_{iN}-\bm{r}_{jB})_{\text{n.n.}}\in\left\{\bm{\delta}^0_1,\bm{\delta}^0_2,\bm{\delta}^0_3\right\}$, where $\bm{\delta}^0_1=-(a_0/\sqrt{3})\bm{e}_y$; $\bm{\delta}^0_2=C_{3z}\bm{\delta}^0_1$; $\bm{\delta}^0_3=C^2_{3z}\bm{\delta}^0_1$. The n.n. FCs are (in $x,y$ basis)
\begin{align}
\Phi_{N,B}(\bm{\delta}^0_1)=\left(\begin{matrix} t_{xx}^{0}& \\ & t_{yy}^{0}\end{matrix} \right), \quad \Phi_{N,B}(\bm{\delta}^0_2)=C_{3z}\Phi_{N,B}(\bm{\delta}^0_1)C_{3z}^{-1},\quad \Phi_{N,B}(\bm{\delta}^0_3)=C^{-1}_{3z}\Phi_{N,B}(\bm{\delta}^0_1)C_{3z}.
\end{align}
The other n.n. FCs are obtained through $\Phi_{B,N}(-\bm{\delta}^0_{j})=\Phi_{N,B}^T(\bm{\delta}^0_j)$ for $j=1,2,3$ ($T$ is the transpose). Each $N$ ($B$) atom has 6 n.n.n. $N$ ($B$) atoms, denoted by 6 relative vectors $(\bm{r}_{iN(B)}-\bm{r}_{jN(B)})_{\text{n.n.n.}}\in\left\{\bm{\delta}^{1}_j (j=1,2,...,6)\right\}$, where $\bm{\delta}^1_1=a_0\bm{e}_x$; $\bm{\delta}^1_{j}=C_{6z}^{j-1}\bm{\delta}^1_1$. The n.n.n. FCs are
\begin{align}
\begin{split}
&\Phi_{N,N}(\bm{\delta}^1_1)=\left(\begin{matrix} t_{xx}^{1}& t_{xy}^1\\ -t_{xy}^1 & t_{yy}^1\end{matrix} \right), \quad \Phi_{N,N}(\bm{\delta}^1_3)=C_{3z}\Phi_{N,N}(\bm{\delta}^1_1)C^{-1}_{3z}, \quad \Phi_{N,N}(\bm{\delta}^1_6)=C^{-1}_{3z}\Phi_{N,N}(\bm{\delta}^1_1)C_{3z}, \\
&\Phi_{N,N}(\bm{\delta}^1_2)=\Phi_{N,N}^T(\bm{\delta}^1_5), \quad 
\Phi_{N,N}(\bm{\delta}^1_4)=\Phi_{N,N}^T(\bm{\delta}^1_1), \quad
\Phi_{N,N}(\bm{\delta}^1_6)=\Phi_{N,N}^T(\bm{\delta}^1_3).
\end{split}
\end{align}
The FCs among $B$ atoms are $\Phi_{B,B}(\bm{\delta}^1_j)=\Phi_{N,N}^T(\bm{\delta}_j^1)$. Notice that $\Phi_{B,B}$ and $\Phi_{N,N}$ are actually independent, i.e., they are not related by any symmetry. In our present simple model, they are set to be related in this way, which is also supported by MD simulations. The onsite FCs are obtained through the sum rule
\begin{align}
\Phi_{N,N}(\bm{0})=-\sum_{j=1}^3\Phi_{N,B}(\bm{\delta}^0_j)-\sum_{j=1}^6\Phi_{N,N}(\bm{\delta}_{j}^1),\quad 
\Phi_{B,B}(\bm{0})=-\sum_{j=1}^3\Phi_{B,N}(\bm{\delta}^0_j)-\sum_{j=1}^6\Phi_{B,B}(\bm{\delta}_{j}^1).
\end{align}
Using these FCs, the dynamical matrix can be easily calculated as
\begin{align}
\begin{split}
& D_{\alpha,\alpha}(\bm{q})=\frac{1}{M_\alpha}\left[\Phi_{\alpha,\alpha}(\bm{0})+\sum_{j=1}^6\Phi_{\alpha,\alpha}(\bm{\delta}^1_j)e^{-i\bm{q}\cdot\bm{\delta}^1_j}\right],\quad \alpha=N,B,\\
& D_{N,B}(\bm{q})=\frac{1}{\sqrt{M_N M_B}}\sum_{j=1}^3\Phi_{N,B}(\bm{\delta}^0_j)e^{-i\bm{q}\cdot\bm{\delta}^0_j}, \quad D_{B,N}(\bm{q})=D_{N,B}^{\dagger}(\bm{q}), \label{eq:D_mono_nonrot_app}
\end{split}
\end{align}
where the mass of atoms is $M_N=14.0067$ amu and $M_B=10.811$ amu. The constants above are found to be (unit: $\text{eV}\cdot\text{\AA}^{-2}$)
\begin{align}
t^0_{xx}=-6.8033,\quad t_{yy}^0=-33.8892,\quad t^1_{xx}=-1.6156,\quad t^1_{xy}=-1.4759, \quad t^1_{yy}=0.2661.
\end{align}
These values are obtain by MD simulations (Sec. \ref{sec_IFC_app}). 

If the lattice is rotated anti-clockwise by $\theta_l=(-1)^l\theta/2$ ($l=1,2$ denote the two layers in twisted bilayer hBN), the intra-layer dynamical matrix $D^0_{l\alpha,l\beta}$ will be rotated from Eq. (\ref{eq:D_mono_nonrot_app}) correspondingly through
\begin{align}
D^0_{l\alpha,l\beta}(\bm{q})=C_{\theta_l}D_{\alpha,\beta}(C^{-1}_{\theta_l}\bm{q})C^{-1}_{\theta_l},\quad C_{\theta_l}=\left(\begin{matrix} \cos\theta_l& -\sin\theta_l\\ \sin\theta_l & \cos\theta_l\end{matrix} \right).
\label{eq:D_mono_analytical_app}
\end{align}

\begin{figure}
\centering
\includegraphics[width=1\textwidth]{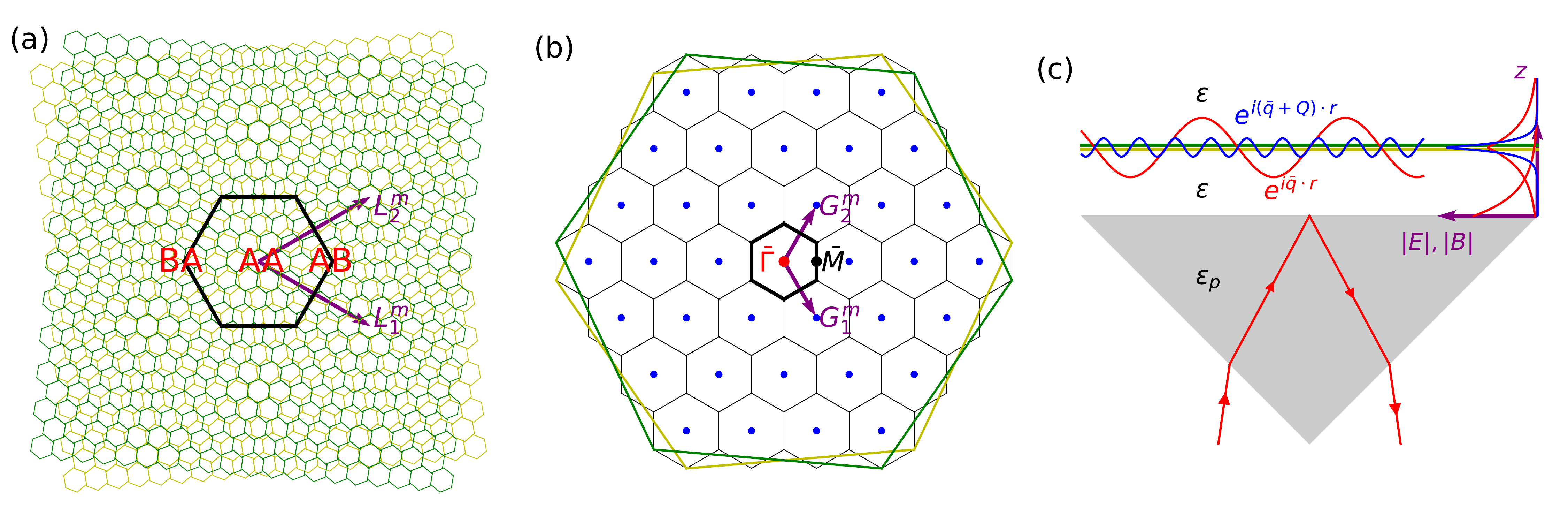}
\caption{\label{fig:fig_setup} (a) The lattice setup of a twisted bilayer hexagonal lattice. The moir\'{e} superlattice is characterized by the supercell (black) and the translation vectors $\bm{L}_{1,2}^m$. Different stacking regions such AA and AB/BA are also formed. (b) The formation of the corresponding mBZ (bold black) and reciprocal basis vectors $\bm{G}_{1,2}^m$. The red dot denotes $\bar{\Gamma}=\bm{0}$. Blue dots denote all other $\bm{Q}\neq\bm{0}$ points, which fold into the $\bar{\Gamma}$ point in the mBZ. For visual clarity, in (a) and (b) we plot the case with $N_a=37$ and $\theta=9.43^\circ$. (c) The Otto configuration for exciting moir\'{e} PhPs in a far-field technique. A prism (gray) is used to generate long-wavelength attenuate incident light (red arrows), which can excite phonons and EM components with shorter wavelengths ($\bm{Q}\neq\bm{0}$, blue).}
\end{figure}

\section{Lattice model of 2D PhP in moir\'{e} polar systems}

\subsection{Equation of motion and polarization in moir\'{e} superlattice}
Moir\'{e} systems have a huge sublattice degree of freedom. Here we study modes of moir\'{e} materials with atomic thickness (not the setup in Ref. \cite{20n_twist_phonon_exp}). We use the notations defined in Refs. \cite{23prb_TAPW,25prb_moire_phonon_tbg}. Suppose each moir\'{e} supercell contains $N_a$ atomic unit cells, i.e., the supercell area $\Omega_m$ and atomic cell area $\Omega_0$ are related by $\Omega_m=N_a\Omega_0$. Each atom's equilibrium position can be labeled as $\bm{r}_{Ii\alpha}=\bm{L}_I+\bm{R}_i+\bm{\tau}_{\alpha}$, where $\bm{L}_I$ is the moir\'{e} Bravais lattice vector, $\bm{R}_i$ (be careful, $i=1,...,N_a$ take finite positions now within a supercell) denotes the atomic Bravais lattice modulo the moir\'{e} lattice, and $\bm{\tau}_{\alpha}$ is the ion position within each atomic cell. We use $\bar{\bm{q}}$ to denote wave vectors in the moir\'{e} Brillouin zone (mBZ), and $\bm{Q}$ the moir\'{e} reciprocal basis vectors (within the atomic Brillouin zone (aBZ), thus there are totally $N_a$ different $\bm{Q}$ vectors \cite{25prb_moire_phonon_tbg}). 

We are interested in the regime with length scale orders
\begin{align}
a_0\ll L_{\theta}\ll\frac{c}{\omega_0}, \label{eq:scale_app}
\end{align}
where $a_0$, $L_{\theta}$ are atomic and supercell lengths, respectively, and $\omega_0$ is the optical phonon frequency. The moir\'{e} potential, here the local force field $\Phi_{i\alpha,j\beta}$, couples different $\bm{Q}$ components (with length scale $L_{\theta}$) together. So we insert a multi-wavelength in-plane field at $z=0$, 
\begin{align}
\bm{E}_t=\sum_{\bm{Q}}\bm{E}_{\bar{\bm{q}}+\bm{Q},t}e^{i(\bar{\bm{q}}+\bm{Q})\cdot\bm{r}-i\omega t},
\end{align}
into the equation of motion of the ionic displacement $\bm{u}$,
\begin{align}
M_{\alpha}\ddot{u}_{\mu}(\bm{r}_{Ii\alpha})+\sum_{Jj\beta\nu}\Phi_{i\alpha\mu,j\beta\nu}(\bm{r}_{Ii\alpha}-\bm{r}_{Jj\beta})u_{\nu}(\bm{r}_{Jj\beta})
-\sum_{\bm{Q}}Z_{\alpha}eE_{\bar{\bm{q}}+\bm{Q},\mu}(\omega)e^{i(\bar{\bm{q}}+\bm{Q})\cdot\bm{r}_{Ii\alpha}-i\omega t}=0. \label{eq:EoM_moire_app}
\end{align}
The polarization density is similarly defined as
\begin{align}
\bm{P}(\bm{r})=\sum_{Ii\alpha}Z_{\alpha}e\bm{u}(\bm{r}_{Ii\alpha})\delta(\bm{r}-\bm{r}_{Ii\alpha}). \label{eq:P_WE_moire_app}
\end{align}
Even in the present moir\'{e} case, the solution is still analytical, and the derivation is almost parallel to the moir\'{e}-less case. Suppose $\bm{u}$ can be expanded as
\begin{align}
\bm{u}(\bm{r}_{Ii\alpha})=\sum_{b}\frac{1}{\sqrt{M_{\alpha}}}e^{i\bar{\bm{q}}\cdot\bm{r}_{Ii\alpha}-i\omega t}\bm{e}_{i\alpha,b}(\bar{\bm{q}})B_b(\bar{\bm{q}},\omega), \label{eq:trial_u_app}
\end{align}
where $\bm{e}_b(\bar{\bm{q}})$ is the eigenvector corresponding to the $b$-th field-free mode with frequency $\omega_b(\bar{\bm{q}})$,
\begin{align}
\sum_{j\beta\nu}D_{i\alpha\mu,j\beta\nu}(\bar{\bm{q}})e_{j\beta\nu,b}(\bar{\bm{q}})=\omega_b^2(\bar{\bm{q}})e_{i\alpha\mu,b}(\bar{\bm{q}}),
\end{align}
and the moir\'{e} dynamical matrix reads
\begin{align}
D_{i\alpha,j\beta}(\bar{\bm{q}})=\sum_{J}\frac{\Phi_{i\alpha,j\beta}(\bm{r}_{Ii\alpha}-\bm{r}_{Jj\beta})}{\sqrt{M_{\alpha}M_{\beta}}}e^{i\bar{\bm{q}}\cdot(\bm{r}_{Jj\beta}-\bm{r}_{Ii\alpha})}. \label{eq:Dmat_moire_app}
\end{align}
The target is to solve for $B_b(\bar{\bm{q}},\omega)$. Plugging Eq. (\ref{eq:trial_u_app}) into Eq. (\ref{eq:EoM_moire_app}) and using the orthogonality relations
\begin{align}
\sum_{i\alpha\mu}e_{i\alpha\mu,b}^*(\bar{\bm{q}})e_{i\alpha\mu,b'}(\bar{\bm{q}})=\delta_{bb'},\quad \sum_{b}e^*_{i\alpha\mu,b}(\bar{\bm{q}})e_{j\beta\nu,b}(\bar{\bm{q}})=\delta_{ij}\delta_{\alpha\beta}\delta_{\mu\nu},
\end{align}
we get
\begin{align}
(\omega_{\bar{\bm{q}}b}^2-\omega^2)B_b(\bar{\bm{q}},\omega)=e\sum_{\bm{Q}}\bm{S}^*_{\bm{Q}b}(\bar{\bm{q}})\cdot\bm{E}_{\bar{\bm{q}}+\bm{Q},t}, \label{eq:a_expression_app}
\end{align}
where the moir\'{e} $S$ matrix has an additional $\bm{Q}$ index,
\begin{align}
\bm{S}_{\bm{Q}b}(\bar{\bm{q}})=\sum_{i\alpha}\frac{Z_{\alpha}\bm{e}_{i\alpha,b}(\bar{\bm{q}})}{\sqrt{M_{\alpha}}}e^{-i\bm{Q}\cdot(\bm{R}_i+\bm{\tau}_{\alpha})}.\label{eq:Smat_app}
\end{align}

We now abstract the continuous field $\bm{P}(\bm{r})$ from Eq. (\ref{eq:P_WE_moire_app}). Writing Eq. (\ref{eq:P_WE_moire_app}) as a Fourier series $\bm{P}_{\bar{\bm{q}}+\bm{Q}}$, but this time we retain some leading terms, i.e., components with $|\bm{Q}|\ll 2\pi/a_0$ ($a_0$ is the atomic cell length; higher-order terms are redundant because they detail information within atomic unit cell)
\begin{align}
\bm{P}(\bm{r})=\sum_{\bm{Q}}\bm{P}_{\bar{\bm{q}}+\bm{Q}}(\omega)e^{i(\bar{\bm{q}}+\bm{Q})\cdot\bm{r}-i\omega t}.
\end{align}
Notice that $\bm{Q}\neq\bm{0}$ terms are necessary to retain here because they reflect the dipole fluctuations at moir\'{e} length scales. The component $\bm{P}_{\bar{\bm{q}}+\bm{Q}}(\omega)$ is calculated as
\begin{align}
\bm{P}_{\bar{\bm{q}}+\bm{Q}}=\frac{1}{N_m\Omega_m}\sum_{Ii\alpha}Z_{\alpha}e\bm{u}(\bm{r}_{Ii\alpha})e^{-i(\bar{\bm{q}}+\bm{Q})\cdot\bm{r}_{Ii\alpha}+i\omega t}
=\frac{e}{\Omega_m}\sum_{b}\bm{S}_{\bm{Q}b}(\bar{\bm{q}})B_b(\bar{\bm{q}},\omega). \label{eq:Pi_qQ_cal_app}
\end{align}
Using Eq. (\ref{eq:a_expression_app}), we can write
\begin{align}
P_{\bar{\bm{q}}+\bm{Q},\mu}(\omega)=\varepsilon_0\sum_{\bm{Q}'\nu}\Pi_{\mu\nu}^{\bm{Q}\bm{Q}'}(\bar{\bm{q}},\omega)E_{\bar{\bm{q}}+\bm{Q}',\nu}, \label{eq:P_expression_app}
\end{align}
where the susceptibility is now a tensor with index $\bm{Q}$,
\begin{align}
\Pi_{\mu\nu}^{\bm{Q}\bm{Q}'}(\bar{\bm{q}},\omega)=\frac{e^2}{\varepsilon_0\Omega_m}\sum_b\frac{[\bm{S}_{\bm{Q}b}(\bar{\bm{q}})]_{\mu}[\bm{S}^*_{\bm{Q}'b}(\bar{\bm{q}})]_{\nu}}{\omega^2_{\bar{\bm{q}}b}-\omega^2}. \label{eq:Pi_result_app} 
\end{align}
It will be convenient to decompose the in-plane fields into directions along and perpendicular to $\bar{\bm{q}}+\bm{Q}$,
\begin{align}
\bm{E}_{\bar{\bm{q}}+\bm{Q},t}=E_{\bar{\bm{q}}+\bm{Q},\parallel}\bm{e}_{\bar{\bm{q}}+\bm{Q},\parallel}+E_{\bar{\bm{q}}+\bm{Q},\perp}\bm{e}_{\bar{\bm{q}}+\bm{Q},\perp},
\end{align}
where 
\begin{align}
\bm{e}_{\bar{\bm{q}}+\bm{Q},\parallel}=(\bar{\bm{q}}+\bm{Q})/|\bar{\bm{q}}+\bm{Q}|,\quad
\bm{e}_{\bar{\bm{q}}+\bm{Q},\perp}=\bm{e}_z\times\bm{e}_{\bar{\bm{q}}+\bm{Q},\parallel}.
\end{align}
Correspondingly, we can decompose $\Pi^{\bm{Q}\bm{Q}'}$ into ($\alpha,\beta=\parallel,\perp$; $\mu,\nu=x,y$)
\begin{align}
\Pi^{\bm{Q}\bm{Q}'}_{\mu\nu}(\bar{\bm{q}},\omega)=\sum_{\alpha\beta}\Pi_{\alpha\beta}^{\bm{Q}\bm{Q}'}(\bar{\bm{q}},\omega)[\bm{e}_{\bar{\bm{q}}+\bm{Q},\alpha}]_{\mu}[\bm{e}_{\bar{\bm{q}}+\bm{Q}',\beta}]_{\nu}.
\end{align}

The moir\'{e} physics enters polaritons by providing the susceptibility $\Pi(\bar{\bm{q}},\omega)$ with off-diagonal $\bm{Q}\neq\bm{Q}'$ terms. In real space, this corresponds to an inhomogeneous optical response. If we turn off the moir\'{e} potential, phonons are simply folded from the moir\'{e}-less system. In other words, $\bm{Q}$ remains a good quantum number, and the mode index $b=(\bm{Q},a)$, where $a$ is the atomic phonon branch index (for hBN, $a=1,...,6$). In this case, $\Pi$ becomes diagonal in $\bm{Q}$, and the diagonal term $\Pi^{\bm{Q}\bm{Q}}(\bar{\bm{q}},\omega)$ reduces to the moir\'{e}-less one with momentum $\bar{\bm{q}}+\bm{Q}$. See subsection \ref{appendix:monolayer_basis_app} for details. 

We note that, in moir\'{e} systems, the optical response is still dominated by the in-plane LO and TO modes. The low-frequency shear and layer-breathing (ZO) modes, while highly sensitive to the moir\'{e} stacking configuration, contribute negligibly to the optical response in the mid-infrared regime. This is because, the dominant atomic displacements in ZO modes are out-of-plane. For a thin 2D sheet, the long-range electric field coupling essential for 2D PhP formation is to the \textit{in-plane} component of the polarization. ZO modes generate a minimal in-plane dipole moment and are therefore not optically active for coupling to in-plane EM waves, unlike the in-plane TO and LO modes. Besides, the energy difference $\hbar|\omega_{\text{TO}} - \omega_{\text{ZO}}|$ is vastly larger than the moir\'{e} potential scattering strength ($\ll 1$ THz). This makes any significant hybridization between high-frequency optical modes and low-frequency ZO modes through the moiré potential physically implausible.

\subsection{Moir\'{e} polaritons}
After the continuation of $\bm{P}$, the whole system of equations is completed by applying Maxwell's equations, using the method in Sec. \ref{subsec:light_incidence_app}. The moir\'{e} material (assumed to have zero thickness) is placed at $z=0$, and the incident light comes from $z<0$. Assume in general that the incident ($l=i$, $z<0$), reflected ($l=r$, $z<0$), and transmitted ($l=t$, $z>0$) light are
\begin{align}
\begin{split}
&\bm{E}^i(\bm{r},t)=\sum_{\bm{Q}\alpha}E^i_{\bar{\bm{q}}+\bm{Q},\alpha}\bm{e}_{\bar{\bm{q}}+\bm{Q},\alpha}e^{i(\bar{\bm{q}}+\bm{Q})\cdot\bm{r}-\lambda_{\bar{\bm{q}}+\bm{Q}}z-i\omega t},\\
&\bm{E}^r(\bm{r},t)=\sum_{\bm{Q}\alpha}E^r_{\bar{\bm{q}}+\bm{Q},\alpha}\bm{e}_{\bar{\bm{q}}+\bm{Q},\alpha}e^{i(\bar{\bm{q}}+\bm{Q})\cdot\bm{r}+\lambda_{\bar{\bm{q}}+\bm{Q}}z-i\omega t},\\
&\bm{E}^t(\bm{r},t)=\sum_{\bm{Q}\alpha}E^t_{\bar{\bm{q}}+\bm{Q},\alpha}\bm{e}_{\bar{\bm{q}}+\bm{Q},\alpha}e^{i(\bar{\bm{q}}+\bm{Q})\cdot\bm{r}-\lambda_{\bar{\bm{q}}+\bm{Q}}z-i\omega t},
\end{split}
\end{align}
where $\alpha$ includes $\parallel,\perp,z$, $\bm{e}_{\bar{\bm{q}}+\bm{Q},z}=\bm{e}_z$, and
\begin{align}
\lambda_{\bar{\bm{q}}+\bm{Q}}= \left\{ \begin{array}{c}
-i\sqrt{\varepsilon\frac{\omega^2}{c^2}-|\bar{\bm{q}}+\bm{Q}|^2}, \quad |\bar{\bm{q}}+\bm{Q}|^2<\varepsilon\frac{\omega^2}{c^2}\\
\sqrt{|\bar{\bm{q}}+\bm{Q}|^2-\varepsilon\frac{\omega^2}{c^2}}, \quad\quad |\bar{\bm{q}}+\bm{Q}|^2>\varepsilon\frac{\omega^2}{c^2}
\end{array}\right.. \label{eq:lambda_z_qQ_app}
\end{align}
This satisfies Eq. (\ref{eq:wave_equation_app}). The divergence law [Eq. (\ref{eq:divergence_theorem_app})] requires
\begin{align}
E_{\bar{\bm{q}}+\bm{Q},z}^i=i\frac{|\bar{\bm{q}}+\bm{Q}|}{\lambda_{\bar{\bm{q}}+\bm{Q}}}E_{\bar{\bm{q}}+\bm{Q},\parallel}^i,\quad
E_{\bar{\bm{q}}+\bm{Q},z}^r=-i\frac{|\bar{\bm{q}}+\bm{Q}|}{\lambda_{\bar{\bm{q}}+\bm{Q}}}E_{\bar{\bm{q}}+\bm{Q},\parallel}^r,\quad
E_{\bar{\bm{q}}+\bm{Q},z}^t=i\frac{|\bar{\bm{q}}+\bm{Q}|}{\lambda_{\bar{\bm{q}}+\bm{Q}}}E_{\bar{\bm{q}}+\bm{Q},\parallel}^t. \label{eq:Ez_app}
\end{align}
The four BCs at the material surface then yield the following relations (expressed in terms of electric fields)
\begin{subequations}
\begin{align}
& \bm{E}^t_{\bar{\bm{q}}+\bm{Q},t} = \bm{E}^i_{\bar{\bm{q}}+\bm{Q},t}+\bm{E}^r_{\bar{\bm{q}}+\bm{Q},t},\\
& E^t_{\bar{\bm{q}}+\bm{Q},z}-E^i_{\bar{\bm{q}}+\bm{Q},z}-E^r_{\bar{\bm{q}}+\bm{Q},z}=-\frac{i(\bar{\bm{q}}+\bm{Q})\cdot \bm{P}_{\bar{\bm{q}}+\bm{Q}}}{\varepsilon_0\varepsilon},\\
& \lambda_{\bar{\bm{q}}+\bm{Q}}(\bm{E}^i_{\bar{\bm{q}}+\bm{Q},t}-\bm{E}^r_{\bar{\bm{q}}+\bm{Q},t}-\bm{E}^t_{\bar{\bm{q}}+\bm{Q},t})-i(\bar{\bm{q}}+\bm{Q})(E^t_{\bar{\bm{q}}+\bm{Q},z}-E^i_{\bar{\bm{q}}+\bm{Q},z}-E^r_{\bar{\bm{q}}+\bm{Q},z})=-\mu_0\omega^2\bm{P}_{\bar{\bm{q}}+\bm{Q}},\\
&\bm{e}_z\cdot\left[(\bar{\bm{q}}+\bm{Q})\times(\bm{E}^t_{\bar{\bm{q}}+\bm{Q},\perp}-\bm{E}^i_{\bar{\bm{q}}+\bm{Q},\perp}-\bm{E}^r_{\bar{\bm{q}}+\bm{Q},\perp})\right]=0.
\end{align}
\end{subequations}
Similar to the toy model case in Sec. \ref{sec:2D_Huang_app}, it is sufficient to consider only the 1st, 2nd, and the transverse components of the 3rd BCs (for each $\bm{Q}$). The 4th BC and the longitudinal component of the 3rd BC coincide with the 1st and 2nd BCs, respectively. Substituting Eqs. (\ref{eq:P_expression_app}) and (\ref{eq:Ez_app}) into the above BCs, and using the 1st BC to eliminate $\bm{E}^r_t$, we can organize these BCs into a set of linear equations
\begin{align}
\sum_{\bm{Q}'\beta}A^{\bm{Q}\bm{Q}'}_{\alpha\beta}(\bar{\bm{q}},\omega)\bm{E}^t_{\bar{\bm{q}}+\bm{Q}',\beta}=\bm{E}^i_{\bar{\bm{q}}+\bm{Q},\alpha}, \label{eq:eigen_equation_moire_app}
\end{align}
with matrix elements
\begin{subequations}
\begin{align}
&A^{\bm{Q}\bm{Q}'}_{\parallel\parallel}(\bar{\bm{q}},\omega)=\delta_{\bm{Q}\bm{Q}'}+\frac{\lambda_{\bar{\bm{q}}+\bm{Q}}}{2\varepsilon}\Pi^{\bm{Q}\bm{Q}'}_{\parallel\parallel}(\bar{\bm{q}},\omega),\\
&A^{\bm{Q}\bm{Q}'}_{\parallel\perp}(\bar{\bm{q}},\omega)= \frac{\lambda_{\bar{\bm{q}}+\bm{Q}}}{2\varepsilon}\Pi^{\bm{Q}\bm{Q}'}_{\parallel\perp}(\bar{\bm{q}},\omega),\\
&A^{\bm{Q}\bm{Q}'}_{\perp\parallel}(\bar{\bm{q}},\omega)=-\frac{1}{2\lambda_{\bar{\bm{q}}+\bm{Q}}}\frac{\omega^2}{c^2}\Pi^{\bm{Q}\bm{Q}'}_{\perp\parallel}(\bar{\bm{q}},\omega),\\
&A^{\bm{Q}\bm{Q}'}_{\perp\perp}(\bar{\bm{q}},\omega)=\delta_{\bm{Q}\bm{Q}'}-\frac{1}{2\lambda_{\bar{\bm{q}}+\bm{Q}}}\frac{\omega^2}{c^2}\Pi^{\bm{Q}\bm{Q}'}_{\perp\perp}(\bar{\bm{q}},\omega).
\end{align}
\end{subequations}
The above equation generalizes Eqs. (\ref{eq:longitudinal_monolayer_app}), (\ref{eq:transverse_monolayer_app}), and is a core result of this paper. From Eq. (\ref{eq:eigen_equation_moire_app}), we can define the transmission and reflection tensors
\begin{align}
T(\bar{\bm{q}},\omega)=A^{-1}(\bar{\bm{q}},\omega), \quad R(\bar{\bm{q}},\omega)=A^{-1}(\bar{\bm{q}},\omega)-I,
\end{align}
such that
\begin{align}
\bm{E}^t_t=T(\bar{\bm{q}},\omega)\bm{E}^i_t, \quad \bm{E}^r_t=R(\bar{\bm{q}},\omega)\bm{E}^i_t,
\end{align}
written in the basis:
\begin{align}
\bm{E}^l_t=(E_{\bar{\bm{q}}+\bm{Q}_1,\parallel}^l,E_{\bar{\bm{q}}+\bm{Q}_1,\perp}^l,...,E_{\bar{\bm{q}}+\bm{Q}_{N_a},\perp}^l)^T.
\end{align}

All information about PhPs is contained in $A(\bar{\bm{q}},\omega)$; for example, the zeros of $\text{det}(A)$ determines the polariton dispersion, and the eigenvectors of $A$ correspond to the respective eigenmode EM fields. In matrix form, Eq. (\ref{eq:eigen_equation_moire_app}) is $A\bm{E}^t=\bm{E}^i$ or $\bm{E}^t=A^{-1}\bm{E}^i$. If the incident light is long-wavelength, i.e., $\bm{E}^{i}=(\bm{E}^{i}_{\bar{\bm{q}}},\bm{0},\bm{0},...)$, then because $A^{-1}$ is not diagonal in $\bm{Q}$ (since $A$ is not diagonal), we generally have $\bm{E}^t_{\bar{\bm{q}}+\bm{Q}}=[A^{-1}]_{\bm{Q}\bm{0}}\bm{E}^i_{\bar{\bm{q}}}\neq \bm{0}$. Physically, this means that a long-wavelength incident light could induce a response with short-wavelength components, via scattering by the moir\'{e} potential (encoded in the $A$ matrix). This effect, depicted in Fig. \ref{fig:fig_setup}(c), is a salient feature of moir\'{e} PhPs, which never occurs in moir\'{e}-less systems.

Based on this property, we can focus on long-wavelength incidence, which already captures information about moir\'{e} scattering and excludes short-wavelength contributions. The effective transmission matrix is the $\bm{Q}=\bm{0}$ submatrix \cite{23sa_plasmon_correlation}:
\begin{align}
T_{\text{eff}}(\bar{\bm{q}},\omega)=[A^{-1}(\bar{\bm{q}},\omega)]^{\bm{0}\bm{0}}. \label{eq:Teff}
\end{align}
The poles of the spectrum $\mathcal{L}(\bar{\bm{q}},\omega)=-\text{Im}[\text{det}[T_{\text{eff}}(\bar{\bm{q}},\omega+i\delta/2)]]$ describe the dispersion of moir\'{e} PhPs that can be excited by long-wavelength light. Since $\mathcal{L}$ can change sign at certain points, we instead plot the spectrum of $\ln(1+|\mathcal{L}(\bar{\bm{q}},\omega)|)$, whose poles are the same to those of $\mathcal{L}$, to visualize the dispersions of moir\'{e} PhPs.

\subsection{Moir\'{e} PhP dispersion against phonon linewidth}

\begin{figure}
\centering
\includegraphics[width=0.6\textwidth]{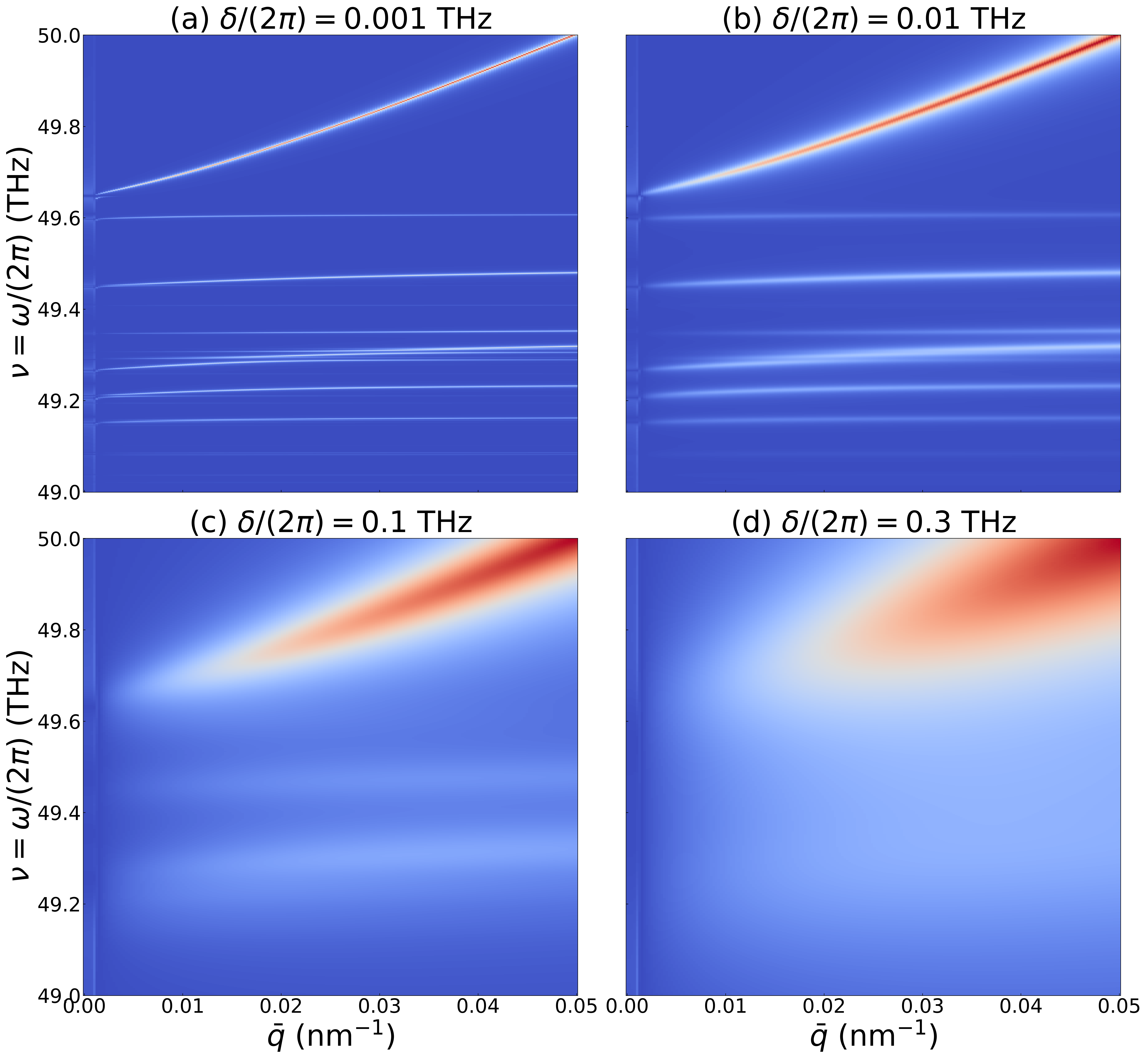}
\caption{\label{fig:fig_against_loss} The PhP dispersion of $2.65^\circ$ twisted bilayer hBN, calculated with the linewidth (a) $\delta/(2\pi)=0.001$ THz, (b) $\delta/(2\pi)=0.01$ THz, (c) $\delta/(2\pi)=0.1$ THz, and (d) $\delta/(2\pi)=0.3$ THz.}
\end{figure}

In Fig. 2 of the main text, the PhP dispersion is plotted using an extremely small phonon linewidth $\delta/(2\pi)=0.001$ THz. Each moir\'{e} branch is distinguishable and separate from the others only if such a tiny loss is assumed. However, in realistic physical systems, the phonon linewidth $\delta$ is finite. It quantifies the finite lifetime of phonons due to intrinsic (e.g., anharmonic scattering, electron-phonon coupling) or external (e.g., defects, boundaries, radiative sources) mechanisms. In general, $\delta$ is momentum- and frequency-dependent, but here we approximate it as constant. Typically, for the optical branches of hBN, the optimized $\delta$ ranges between $0.2$-$0.5$ THz \cite{18prb_hBN_linewidth,18nm_hBN_linewidth,20prl_hBN_linewidth}, which is much larger than the value adopted in Fig. 2 of the main text. 

To determine whether the moir\'{e} PhP dispersion can survive under realistic conditions, we calculate the PhP spectrum of twisted bilayer $2.65^\circ$ hBN for different phonon linewidths: $0.001$, $0.01$, $0.1$, and $0.3$ THz. The results are shown in Fig. \ref{fig:fig_against_loss}. The dispersion is clearly visible for $\delta/(2\pi)\leq 0.01$ THz [Fig. \ref{fig:fig_against_loss}(a)(b)], and becomes obscure as $\delta$ increases further. In the range of realistic phonon linewidth ($\delta\geq0.1$ THz), some moir\'{e} branches `merge' with others, and only the rough outlines are recognizable. For $\delta/(2\pi)=0.1$ THz [Fig. \ref{fig:fig_against_loss}(c)], we can still distinguish some moir\'{e} branches, though the fine structure shown in Fig. 2(b) of the main text is obscured. For $\delta/(2\pi)=0.3$ THz [Fig. \ref{fig:fig_against_loss}(d)], the spectrum becomes more mixed. In this case, we can hardly distinguish the dispersion of each PhP branch from the spectrum, but we can roughly identify the frequency region ($49.1$-$49.5$ THz) where moir\'{e} physics dominates. 

We find that the PhP spectrum depends quite sensitively on $\delta$ in the realistic range. The moir\'{e} PhP dispersion can be captured only in extremely optimized samples with sufficiently low loss (e.g., $\delta\approx 0.1$ THz). Thus, the direct detection of PhP dispersion using traditional far-field setups is challenging. Instead, as shown in Figs. 3 and 4 of the main text, electric fields fluctuate in real space at specific frequencies, proving a more compelling signal to verify the existence of moir\'{e} PhPs.

\subsection{PhP spectrum of twisted bilayer MoTe2 system}
In MoTe$_2$, the ions carry effective charges $Z_{\text{Mo}}=-2Z_{\text{Te}}\approx 3.16$ (in units of $e$, obtained from DFT calculations), and the ionic masses are larger than those of hBN. The critical optical phonon frequency is $\omega_0/(2\pi)\approx 7.2$ THz. For twisted bilayer systems, the force constant model can be constructed using MD simulations (Section \ref{sec_IFC_app}). 

\begin{figure}
\centering
\includegraphics[width=1\textwidth]{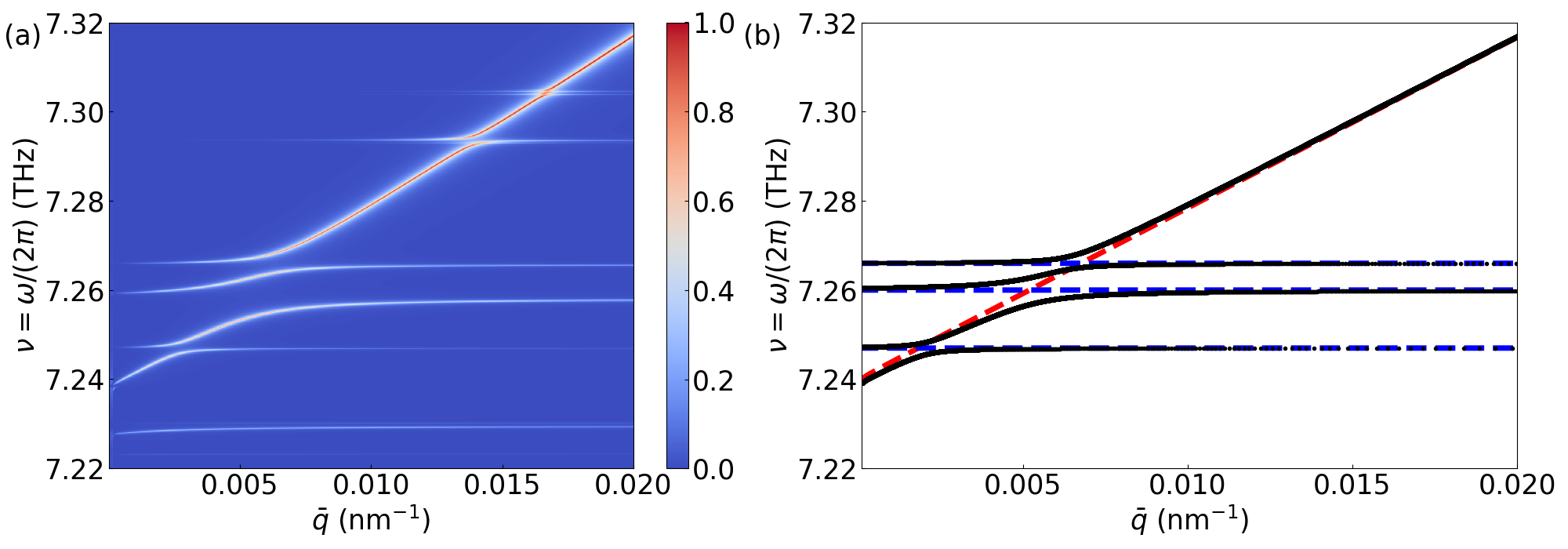}
\caption{\label{fig:fig_MoTe2} (a) Long-wavelength PhP dispersion (along the $\bar{\Gamma}-\bar{M}$ direction) for $3.89^{\circ}$ twisted bilayer MoTe$_2$, near the frequency $\nu_0\approx 7.2$ THz ($q_0\approx 1.5\times10^{-4}$ nm$^{-1}$), calculated from the lattice model. Here we take $\delta/2\pi=2\times10^{-4}$ THz. (b) The same PhP dispersion (black) obtained using the coupled oscillator model (TM mode). Here we take $\gamma_0^2/(2\pi)^4=3.9\times10^7$ CV$^{-1}$s$^{-2}$, and for $\mathcal{D}$ parameters [in the unit of $(2\pi \text{THz})^2$]: $\mathcal{D}_{00}=7.240^2$, $\mathcal{D}_{11}=7.247^2$, $\mathcal{D}_{22}=7.260^2$, $\mathcal{D}_{33}=7.266^2$, $\mathcal{D}_{01}=\mathcal{D}_{10}=0.0204$, $\mathcal{D}_{02}=\mathcal{D}_{20}=0.0456$, $\mathcal{D}_{03}=\mathcal{D}_{30}=0.0236$, (all other parameters are set to zero), obtained by fitting (a). For comparison, we also show the PhP dispersion (red dashed line) and the high-momentum $\bm{Q}$ phonon frequencies (blue dashed lines) for the decoupled case.}
\end{figure}

We have calculated the PhP spectrum of $3.89^{\circ}$ twisted bilayer MoTe$_2$, which has aroused wide interest due to the discovery of its fractional Chern insulating phase. The result is shown in Fig. \ref{fig:fig_MoTe2}(a). Comparing it with hBN (Fig. 2 of the main text), we find the qualitative structures of the PhP dispersion are the same. For example, the PhP dispersions consist of some mini-branches and a dominant branch. The mini-branches have smaller bandwidths and intensities, encoding mainly moir\'{e} components, while the dominant branch has a linear slope and stronger intensity, inheriting mainly from the moir\'{e}-less TM mode. Additionally, the intensities of the mini-branches weaken as $\omega$ moves far from $\omega_0$.

We then analyze some differences with hBN, most of which are quantitative. Both the critical frequency $\omega_0$ and the mini-band gaps (at $\bar{\bm{q}}=\bm{0}$, $\approx0.01$ THz) are smaller than those in hBN. This is mainly due to the heavier mass in MoTe$_2$. The number of mini-branches is smaller than in hBN, because the $3.89^\circ$ system has a much smaller supercell size than the $2.65^\circ$ supercell in hBN. More moir\'{e} PhP branches (and smaller band gaps) are expected with a smaller twisting angle. 

\section{Macroscopic theory of moir\'{e} PhP \label{sec:append_coupled_oscillator_app}}

\subsection{A toy model: coupled harmonic oscillators}
In the presence of inhomogeneity, the system is described by a set of coupled vibration fields $\bm{W}_a$ and electric fields $\bm{E}_a$. In the present case $a$ labels just the different Fourier components $\bm{Q}$. More specifically, we use $\bm{W}_0$ and $\bm{E}_0$ to denote the usual long-wavelength fields (with momentum $\bar{\bm{q}}\approx\bm{0}$), while all other components ($a\neq 0$) correspond to high-momentum fields (with momentum $\bar{\bm{q}}+\bm{Q}$). The equation of motion and polarization density for the $a$-th component are
\begin{align}
\ddot{\bm{W}}_a=-\sum_{a'}\mathcal{D}_{aa'}\bm{W}_{a'}+\gamma\bm{E}_a,\quad
\bm{P}_a=\gamma\bm{W}_a, \label{eq:B1_app}
\end{align}
where $\mathcal{D}_{aa'}=D^*_{a'a}$ is the coupling constant among $\bm{W}_a$ and $\bm{W}_{a'}$, $\gamma$ is the charge coefficient. Notice that $\mathcal{D}_{aa}=\omega^2_a$ is the elastic eigenfrequency. The nontrivial inhomogeneity leads to $\mathcal{D}_{aa'}\neq 0$ when $a\neq a'$. We can decompose $\bm{W}_a$ into normal modes $\bm{W}_a=\sum_{b}U_{ab}\bm{w}_{b}$ that diagonalize the coupled system, where the orthogonal transformation matrix $V$ satisfies
\begin{align}
\sum_{a}(\omega^2_b\delta_{a'a}-\mathcal{D}_{a'a})U_{ab}=0, \quad
\sum_{a}U^*_{ab}U_{ab'}=\delta_{bb'},\quad \sum_{b}U_{ab}U^*_{a'b}=\delta_{aa'}.
\end{align}
One can show that (for oscillation at frequency $\omega$)
\begin{align}
\ddot{\bm{w}}_b=-\omega_b^2\bm{w}_b+\gamma\sum_{a}U^*_{ab}\bm{E}_a,\quad
\bm{P}_a=\gamma\sum_{b}U_{ab}\bm{w}_{b}=\sum_{a'}\sum_{b}\gamma^2\frac{U_{ab}U^*_{a'b}}{\omega_b^2-\omega^2}\bm{E}_{a'}.
\end{align}
If we focus on the long-wavelength dispersion, we may neglect all short-wavelength fields $\bm{w}_b$, $\bm{E}_b$, and $\bm{P}_b$ with $b\neq 0$. If so, after redefining $T_b=|\gamma U_{0b}|^2/\varepsilon_0$, we obtain the effective long-wavelength susceptibility
\begin{align}
\Pi(\omega)=\Pi^{00}(\omega)=\sum_{b}\frac{T_b}{\omega_b^2-\omega^2},
\end{align}
which has multiple poles. By plugging this long-wavelength susceptibility into Eq. (\ref{eq:TM_condition_app}), we can obtain the moir\'{e} PhP dispersion. In Fig. \ref{fig:fig_MoTe2} we plot the TM dispersion using this method and fitting the model of twisted MoTe$_2$. Notice that each pole $\omega_b$ has a specific strength $T_b$, proportional to the scattering strength into that channel, and gives rise to a specific pair of TM and TE modes in the absence of other poles. The TE modes are mainly lattice vibrations, contributing little to the PhP dispersion. The TM modes disperse linearly with slopes proportional to $T_b$ starting from $\omega_b$ [Eq. (\ref{eq:LO_dispersion_app})]. Therefore, crossings occur between these branches with different slopes if they are not coupled. The whole frequency region is partitioned into a series of negative windows where $\Pi(\omega)<0$ and positive windows where $\Pi(\omega)>0$. Remembering the sign rule, the TM (TE) PhP lies separately in the negative (positive) window. This is true even when all the poles are taken into consideration. This results in the anti-crossings shown in Fig. \ref{fig:fig_MoTe2}(b). The most dispersive (dominant) branch has the largest $T_b$, which carries the most long-wavelength component. The moir\'{e} pattern of $\bm{E}$ and $\bm{W}$ is a byproduct: when $\omega$ approaches $\omega_b$, the lattice oscillates exclusively in accordance with mode $b$, which generates a short-wavelength polarization, thus jointly inducing the moir\'{e} EM fields. 

We observe that Ref. \cite{21n_exciton_polariton_exp} adopts a similar model to explain the exciton polariton. Their two poles originate from the heterostructure. Here, the multiple poles result from umklapp scattering due to the inhomogeneous moir\'{e} potential in the homobilayer. This is also why these poles $\omega_b$ are so close to each other. The coupled oscillator model presented in this subsection is only a toy model. It is only used to illustrate how the moir\'{e} potential produces multiple branches of PhP. We will refine this model in the next subsection to make it more quantitative and accurate.

\subsection{The continuum model for moir\'{e} phonon \label{sec:continuum_phonon_app}}
A more accurate model should incorporate the anisotropy of TO and LO phonons at various moir\'{e} reciprocal vectors $\bm{Q}$. The macroscopic model for moir\'{e} PhP requires a corresponding continuum model for optical moir\'{e} phonons (without long-wavelength electric field). The theoretical structure is quite similar to the Bistritzer-MacDonald (BM) model for electrons in the magic-angle twisted bilayer graphene \cite{11pnas_BM_model}, and is more conveniently obtained using the truncated plane atomic wave (TAPW) method \cite{23prb_TAPW,25prb_moire_phonon_tbg}.

\begin{figure}
\centering
\includegraphics[width=1\textwidth]{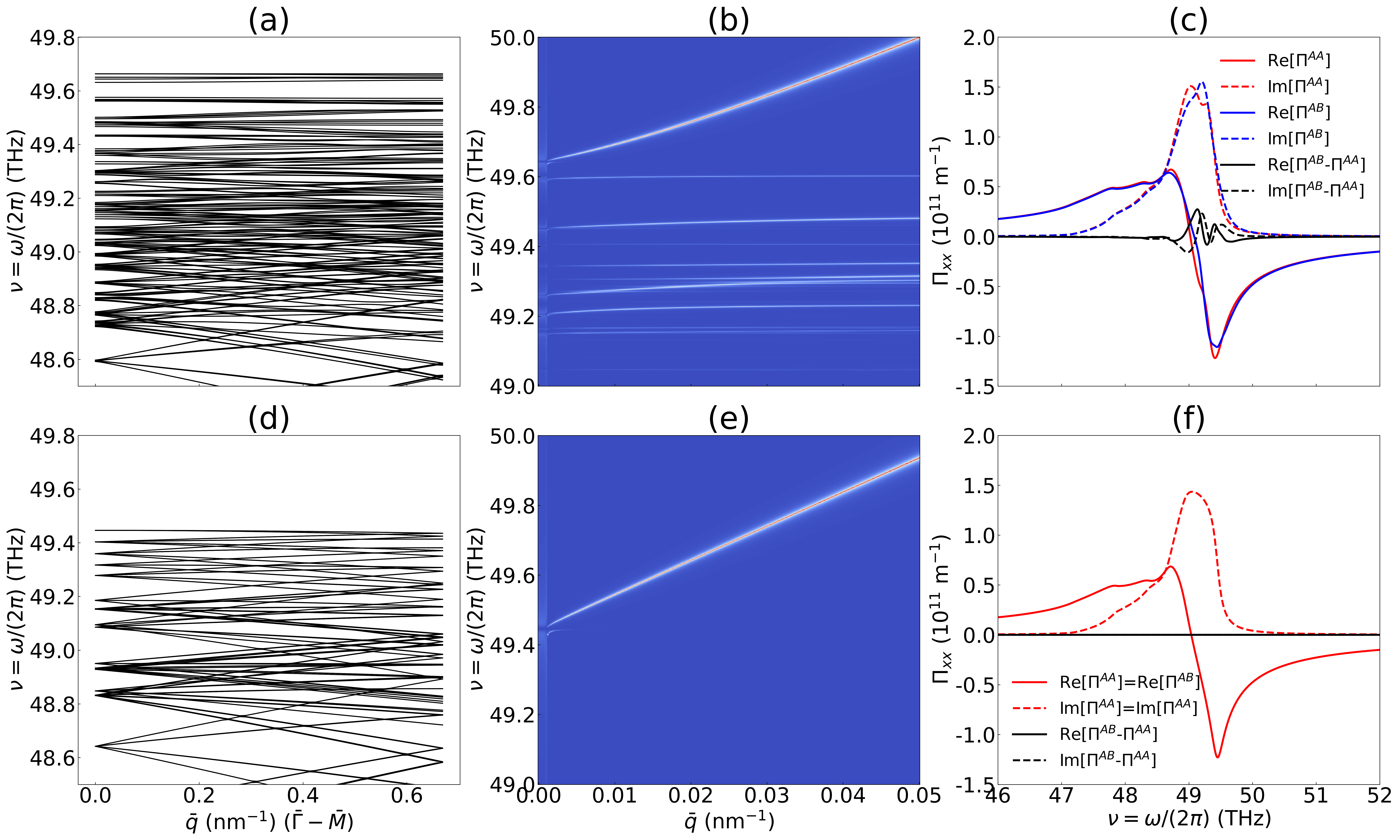}
\caption{\label{fig:fig_moire_vs_decoupled} Phonon bands [(a), (d)], PhP bands [(b), (e)], and the local susceptibility $\Pi_{xx}(\bm{r},\bm{r},\omega)$ [(c), (f)], calculated using the continuum model from Secs. \ref{sec:continuum_phonon_app} and \ref{sec:continuum_polariton_app}. (a), (b), (c) show results for $2.65^\circ$ twisted bilayer hBN, while in (d), (e), (f) we have turned off the moir\'{e} potential, i.e., we set $\delta\mathcal{D}^{\text{intra}}=\delta\mathcal{D}^{\text{inter}}=0$. All calculations use a basis of $61$ $\bm{Q}$ vectors. In (b), (e) we use the linewidth $\delta/(2\pi)=0.001$ THz. In (c), (f) we use $\delta/(2\pi)=0.15$ THz.}
\end{figure}

The dynamical matrix (\ref{eq:Dmat_moire_app}) has exactly the same form as the tight-binding Hamiltonian for electrons, thus suggesting the same mathematical and physical structure. The general idea of the TAPW method is to expand the eigenmodes of the moir\'{e} system using the monolayer ``plane waves'', i.e., 
\begin{align}
e_{i \alpha \mu, b} (\bar{\bm{k}})=\sum_{\bm{Q}la} \frac{e^{i\bm{Q}\cdot(\bm{R}_{i}+\bm{\tau}_{\alpha})}}{\sqrt{N_a}}e_{\alpha\mu,la}(\bar{\bm{q}}+\bm{Q})U_{\bm{Q}la,b} (\bar{\bm{q}}), \label{eq:band_transform_app}
\end{align}
where the vector $\bm{e}_{la}(\bm{q})$ ($a$: branch, $l$: layer) satisfies the monolayer dynamical equation Eq. (\ref{eq:moireless_EoM_app}) and orthogonality Eq. (\ref{eq:moireless_orthogonality_app}). If we turn off the moir\'{e} potential, $U(\bar{\bm{q}})$ becomes diagonal and the phonons simply reduce to those folded from the monolayer case without any hybridization (Sec. \ref{appendix:monolayer_basis_app}). We can expect that a small truncation of $\bm{Q}$ vectors in the $\Gamma$ valley (here we take $N_{\bm{Q}}=61$) is sufficient to calculate the phonons there. Since we only focus on the iLO and iTO modes, we limit the summation over $a$ to these two branches. Therefore, the TAPW method reduces the dimension to $4N_{\bm{Q}}$. The $U$ matrix satisfies
\begin{align}
\sum_{\bm{Q}'l'a'}\mathcal{D}_{\bm{Q}la,\bm{Q}'l'a'}(\bar{\bm{q}})U_{\bm{Q}'l'a',b}(\bar{\bm{q}})=\omega^2_{\bar{\bm{q}}b}U_{\bm{Q}la,b}(\bar{\bm{q}}),
\end{align}
where the transformed dynamical matrix is derived to be \cite{23prb_TAPW,25prb_moire_phonon_tbg}
\begin{align}
\mathcal{D}_{\bm{Q}la,\bm{Q}'l'a'}(\bar{\bm{q}})=\sum_{i\alpha\mu,j\beta\nu} e^*_{\alpha\mu,la}(\bar{\bm{q}}+\bm{Q})\frac{e^{- i\bm{Q} \cdot (\bm{R}_{i}+\bm{\tau}_{\alpha})}}{\sqrt{N_a}} D_{i\alpha\mu, j\beta\nu} (\bar{\bm{q}})
\frac{e^{i\bm{Q}'\cdot(\bm{R}_{j}+\bm{\tau}_{\beta})}}{\sqrt{N_a}}e_{\beta\nu,l'a'}(\bar{\bm{q}}+\bm{Q}'). 
\end{align}

Our next goal is to simplify $\mathcal{D}(\bar{\bm{q}})$. For convenience, we fix the direction of iLO/iTO basis vectors at different $\bar{\bm{q}}+\bm{Q}$, and use the following basis at $\bm{q}=\bm{0}$ to replace $\bm{e}_{\alpha,la}(\bm{q})$ [see Eq. (\ref{eq:LOTO_at_0_express_app})]
\begin{align}
\bm{e}_{lx}=\left(\sqrt{\frac{M_B}{M_N+M_B}},0,-\sqrt{\frac{M_N}{M_N+M_B}},0\right)^T,\quad
\bm{e}_{ly}=\left(0,\sqrt{\frac{M_B}{M_N+M_B}},0,-\sqrt{\frac{M_N}{M_N+M_B}}\right)^T. \label{eq:xy_express_app}
\end{align}
This step itself is an approximation that assumes the Hilbert space spanned by iLO/iTO modes at $\bar{\bm{q}}+\bm{Q}$ is the same as the $\Gamma$ point. But it greatly simplifies the calculation since it avoids the complicated $\bm{q}$-dependence of $\bm{e}_{la}(\bm{q})$ in the spirit of $k\cdot p$ theory. Correspondingly, we replace the branch index $a=$iLO,iTO with the index $\mu=x,y$. Note that here $\mu$ indexes the mode, not the spatial direction. The dynamical matrix in the basis of Eq. (\ref{eq:xy_express_app}) then becomes
\begin{align}
\mathcal{D}_{\bm{Q}l\mu,\bm{Q}'l'\nu}(\bar{\bm{q}})=\sum_{i\alpha,j\beta}\frac{e^{- i\bm{Q} \cdot (\bm{R}_{i}+\bm{\tau}_{\alpha})}}{\sqrt{N_a}} \bm{e}^*_{\alpha,l\mu}D_{i\alpha, j\beta} (\bar{\bm{q}})\bm{e}_{\beta,l'\nu}
  \frac{e^{i\bm{Q}'\cdot(\bm{R}_{j}+\bm{\tau}_{\beta})}}{\sqrt{N_a}}.
\end{align}
The intralayer $l=l'$ terms with $\bm{Q}=\bm{Q}'$ are straightforward to evaluate. We can simply diagonalize the monolayer dynamical matrix at $\bar{\bm{q}}+\bm{Q}$ [i.e., $D^0_{l\alpha,l\beta}(\bar{\bm{q}}+\bm{Q})$ defined in Eq. (\ref{eq:D_mono_analytical_app})], and use the monolayer iLO/iTO frequencies (denoted by $\omega^0_{\bar{\bm{q}}+\bm{Q},la}$) to write
\begin{align}
\mathcal{D}^0_{\bm{Q}l\mu,\bm{Q}l\nu}(\bar{\bm{q}})=\sum_{a}^{\text{iLO,iTO}}(\omega^0_{\bar{\bm{q}}+\bm{Q},la})^2\bm{e}_{l\mu}^{T}\bm{e}_{la}(\bar{\bm{q}}+\bm{Q})\bm{e}^T_{la}(\bar{\bm{q}}+\bm{Q})\bm{e}_{l\nu}, \label{eq:Dmat_mono_app}
\end{align}
where $\bm{e}_{la}(\bm{q})$ is defined in Eq. (\ref{eq:LOTO_express_app}). We now introduce the moir\'{e} potential. The moir\'{e} potential generally consists of two parts: an intralayer part, arising primarily from lattice relaxation, and an interlayer part, due mainly to commensurate interlayer scattering. For simplicity, we retain only the hoppings among nearest $\bm{Q}$ vectors (just like the BM model). The intralayer moir\'{e} potential can then be expressed in the $\bar{\bm{q}}$-independent form
\begin{align}
\delta\mathcal{D}^{\text{intra}}_{\bm{Q}l\mu,\bm{Q}'l\nu} = \sum_{j=1,2,3}\left[B_{l\mu,l\nu}(\bm{G}_j^m)\delta_{\bm{Q},\bm{Q}'+\bm{G}_j^m}+B^*_{l\mu,l\nu}(\bm{G}_j^m)\delta_{\bm{Q},\bm{Q}'-\bm{G}_j^m} \right], \label{eq:Dmat_intra_app}
\end{align}
where $\bm{G}_1^m=4\pi/(\sqrt{3}L_{\theta})(1/2,-\sqrt{3}/2)$, $\bm{G}_2^m=C_{3z}\bm{G}_1^m$, $\bm{G}_3^m=C_{3z}^{-1}\bm{G}_1^m$, $L_{\theta}$ is the moir\'{e} supercell length. The $B$ matrix with argument $\bm{G}_1^m$ is found to be
\begin{align}
B_{1,1}(\bm{G}_1^m)
=\left(\begin{matrix} 256.20-3.09i & -161.46-4.68i \\-161.46-4.68i & -292.64-2.84i\end{matrix}\right), \label{eq:B_11_app}
\end{align}
for $2.65^\circ$ bilayer hBN. The numbers here (and below) are in the unit of THz$^2$. The interlayer part is found to be much smaller than the intralayer part. It is given by ($\bar{l}$ denotes another layer of $l$)
\begin{align}
\delta\mathcal{D}^{\text{inter}}_{\bm{Q}l\mu,\bm{Q}'\bar{l}\nu} =B_{l\mu,\bar{l}\nu}(\bm{0})\delta_{\bm{Q},\bm{Q}'}+\sum_{j=1,2,3}\left[B_{l\mu,\bar{l}\nu}(\bm{G}_j^m)\delta_{\bm{Q},\bm{Q}'+\bm{G}_j^m} + B^*_{l\mu,\bar{l}\nu}(\bm{G}_j^m)\delta_{\bm{Q},\bm{Q}'-\bm{G}_j^m}\right], \label{eq:Dmat_inter_app}
\end{align}
with
\begin{align}
B_{1,2}(\bm{0})
=\left(\begin{matrix} -7.64 &  \\ & -7.64 \end{matrix}\right),\quad
B_{1,2}(\bm{G}_1^m)
=\left(\begin{matrix} 9.20 & 6.74 \\ 6.74 & 1.41 \end{matrix}\right).
\end{align}
The other $B$ matrices can be obtained through time reversal $\mathcal{T}$, $C_{3z}$ and $C_{2y}$ rotations:
\begin{align}
\begin{split}
&B_{\bar{l},\bar{l'}}(\bm{G}_1^m)=C_{2y}C_{3z}B^*_{l,l'}(\bm{G}_1^m)C_{3z}^{T}C_{2y},\\ 
&B_{l,l'}(\bm{G}_2^m)=C_{3z}B_{l,l'}(\bm{G}_1^m)C_{3z}^{T},\\ 
&B_{l,l'}(\bm{G}_3^m)=C_{3z}^{T}B_{l,l'}(\bm{G}_1^m)C_{3z}.
\end{split}
\end{align}

By treating the phonon fields as continuum plane waves indexed by layer and sublattice, the total dynamical matrix can be written in a more compact form that resembles the BM Hamiltonian, 
\begin{align}
D(-i\nabla,\bm{r})=\left( \begin{matrix} \mathcal{D}^0_{1}(-i\nabla) + \delta\mathcal{D}_{1,1}^{\text{intra}}(\bm{r}) & \delta\mathcal{D}_{1,2}^{\text{inter}}(\bm{r}) \\ \delta\mathcal{D}_{2,1}^{\text{inter}}(\bm{r}) & \mathcal{D}_{2}^0(-i\nabla)+ \delta\mathcal{D}_{2,2}^{\text{intra}}(\bm{r}) \end{matrix} \right),
\end{align}
where $\mathcal{D}^0_l(-i\nabla)$ represents the iLO/iTO frequencies and is diagonal in $\bm{Q}$ with matrix element given by Eq. (\ref{eq:Dmat_mono_app}), and
\begin{align}
\begin{split}
&\delta\mathcal{D}^{\text{intra}}_{l,l}(\bm{r}) = \sum_{j=1,2,3}B_{l,l}(\bm{G}_j^m)e^{i\bm{G}_j^m\cdot\bm{r}} + h.c.,\\
&\delta\mathcal{D}^{\text{inter}}_{l,\bar{l}}(\bm{r}) =\frac{1}{2}B_{lA,\bar{l}A'}(\bm{0})+\sum_{j=1,2,3}B_{l,\bar{l}}(\bm{G}_j^m)e^{i\bm{G}_j^m\cdot\bm{r}} + h.c.. \label{eq:moire_potential_app}
\end{split}
\end{align}

\subsection{The continuum model for moir\'{e} PhP \label{sec:continuum_polariton_app}}
The continuum model for moir\'{e} PhP can be easily obtained through the moir\'{e} phonon model introduced in the last subsection. Firstly, we notice that in moir\'{e} systems, the continuum version of displacement field $\bm{u}_{l\alpha}$ can be written as
\begin{align}
\bm{u}_{l\alpha}(\bm{r},t) = \sum_{\bm{Q}}\bm{u}_{l\alpha,\bar{\bm{q}}+\bm{Q}}(\bm{r},t),
\end{align}
where $\bm{u}_{l\alpha,\bar{\bm{q}}+\bm{Q}}\propto e^{i(\bar{\bm{q}}+\bm{Q})\cdot\bm{r}}$ is the $\bm{Q}$-th Fourier component of $\bm{u}_{l\alpha}$ in layer $l$. Like Eq. (\ref{eq:W_microscopic_app}), we define a series of continuum fields, characterized by the wavevector $\bm{Q}$, at layer $l$, 
\begin{align}
\bm{W}_{\bar{\bm{q}}+\bm{Q},l}(\bm{r},t)=\frac{1}{\sqrt{\Omega_0}}\sqrt{\frac{M_NM_B}{M_N+M_B}}[\bm{u}_{lN,\bar{\bm{q}}+\bm{Q}}(\bm{r},t)-\bm{u}_{lB,\bar{\bm{q}}+\bm{Q}}(\bm{r},t)]=\bm{W}_{\bar{\bm{q}}+\bm{Q},l}e^{i(\bar{\bm{q}}+\bm{Q})\cdot\bm{r}}.
\end{align}
The complete continuum $\bm{W}$ field consists of components with different wavevectors and layers: $\bm{W}(\bm{r},t)=\sum_{\bm{Q}l}\bm{W}_{\bar{\bm{q}}+\bm{Q},l}e^{i(\bar{\bm{q}}+\bm{Q})\cdot\bm{r}}$, so does the continuum $\bm{P}$ field:
\begin{align}
\bm{P}(\bm{r},t) = \frac{1}{\Omega_0}\sum_{l\alpha}Z_{\alpha}e\bm{u}_{l\alpha}(\bm{r},t)=\gamma\sum_{\bm{Q}l}\bm{W}_{\bar{\bm{q}}+\bm{Q},l}e^{i(\bar{\bm{q}}+\bm{Q})\cdot\bm{r}}=\sum_{\bm{Q}}\bm{P}_{\bar{\bm{q}}+\bm{Q}}e^{i(\bar{\bm{q}}+\bm{Q})\cdot\bm{r}},
\end{align}
where $\gamma=Z_Ne(M_N^{-1}+M_B^{-1})^{1/2}\Omega_0^{-1/2}$ takes the same form as the monolayer case in Eq. (\ref{eq:gamma_microscopic_app}). We see that the $\bm{Q}$-component of $\bm{P}$ and $\bm{W}$ are related simply through
\begin{align}
\bm{P}_{\bar{\bm{q}}+\bm{Q}}=\gamma(\bm{W}_{\bar{\bm{q}}+\bm{Q},1}+\bm{W}_{\bar{\bm{q}}+\bm{Q},2}).
\end{align}
With the multi-component electric field $\bm{E}(\bm{r},t)=\sum_{\bm{Q}}\bm{E}_{\bar{\bm{q}}+\bm{Q}}e^{i(\bar{\bm{q}}+\bm{Q})\cdot\bm{r}}$, the equation of motion for $\bm{W}$ becomes a hybrid matrix equation
\begin{align}
\ddot{W}_{\bar{\bm{q}}+\bm{Q},l\mu}=-\sum_{\bm{Q}'l'\nu}\mathcal{D}_{\bm{Q}l\mu,\bm{Q}'l'\nu}(\bar{\bm{q}})W_{\bar{\bm{q}}+\bm{Q}',l'\nu}+\gamma E_{\bar{\bm{q}}+\bm{Q},\mu},
\end{align}
where $\mathcal{D}(\bar{\bm{q}})$ is simply the dynamical matrix in the continuum model of moir\'{e} phonons, under the basis in Eq. (\ref{eq:xy_express_app}), i.e., Eqs. (\ref{eq:Dmat_mono_app}), (\ref{eq:Dmat_intra_app}), and (\ref{eq:Dmat_inter_app}). For further simplification, we can even neglect the $\bar{\bm{q}}$-dependence of the force field, i.e., $\mathcal{D}(\bar{\bm{q}})=\mathcal{D}(\bar{\bm{0}})$. If so, the $\bar{\bm{q}}$-dependence of the PhP only comes from the EM waves, i.e., through the introduction of parameters $\lambda_{\bm{Q}}^2=(\bar{\bm{q}}+\bm{Q})^2-\omega^2/c^2$. For a more accurate calculation, we should retain the $\bar{\bm{q}}$-dependence of the force field.

The following discussion is parallel to the toy model. Suppose $\bm{w}_{\bar{\bm{q}}b}$ diagonalizes $\mathcal{D}(\bar{\bm{q}})$, and the plane waves are expanded using normal modes as $W_{\bar{\bm{q}}+\bm{Q},l\mu}=\sum_{b}U_{\bm{Q}l\mu,b}(\bar{\bm{q}})w_{\bar{\bm{q}},b}$ [notice that $\mathcal{D}(\bar{\bm{q}})U(\bar{\bm{q}})=U(\bar{\bm{q}})\text{Diag}(\omega_{\bar{\bm{q}}b}^2)$]. Then, if the system is driven by an electric field at frequency $\omega$, we can solve
\begin{align}
w_{\bar{\bm{q}}b}=\frac{\gamma}{\omega_{\bar{\bm{q}}b}^2-\omega^2}\sum_{\bm{Q}l\mu}E_{\bar{\bm{q}}+\bm{Q},\mu}U^*_{\bm{Q}l\mu,b}(\bar{\bm{q}}).
\end{align}
So the polarization
\begin{align}
P_{\bar{\bm{q}}+\bm{Q},\mu}=\gamma\sum_{l}W_{\bar{\bm{q}}+\bm{Q},l\mu}=\gamma\sum_{lb}w_{\bar{\bm{q}}b}U_{\bm{Q}l\mu,b}(\bar{\bm{q}})=\sum_{\bm{Q}'\nu}\sum_{bll'}\frac{\gamma^2}{\omega_{\bar{\bm{q}}b}^2-\omega^2}U_{\bm{Q}l\mu,b}(\bar{\bm{q}})U^*_{\bm{Q}'l'\nu,b}(\bar{\bm{q}})E_{\bar{\bm{q}}+\bm{Q}',\nu},
\end{align}
from which we read
\begin{align}
\varepsilon_0\Pi_{\mu\nu}^{\bm{Q}\bm{Q}'}(\bar{\bm{q}}) = \gamma^2\sum_{bll'}\frac{U_{\bm{Q}l\mu,b}(\bar{\bm{q}})U^*_{\bm{Q}'l'\nu,b}(\bar{\bm{q}})}{\omega_{\bar{\bm{q}}b}^2-\omega^2}. \label{eq:Pi_moire_continuum_app}
\end{align}

We have listed some results in Fig. \ref{fig:fig_moire_vs_decoupled} calculated using the present continuum model for the $2.65^\circ$ twisted bilayer hBN. Both the moir\'{e} PhP bands and local susceptibility are accurately recovered [compared to those shown in the main text obtained through the lattice model]. We have tried artificially turned off the moir\'{e} potential by setting Eq. (\ref{eq:moire_potential_app}) to zero. In such a case, the system consists of two decoupled monolayers, where all moir\'{e} physics disappear: the multiple flat moir\'{e} PhP bands are missing, and the local susceptibility shows no signal difference between AA- and AB-stacking points.

\begin{figure}
\centering
\includegraphics[width=0.6\textwidth]{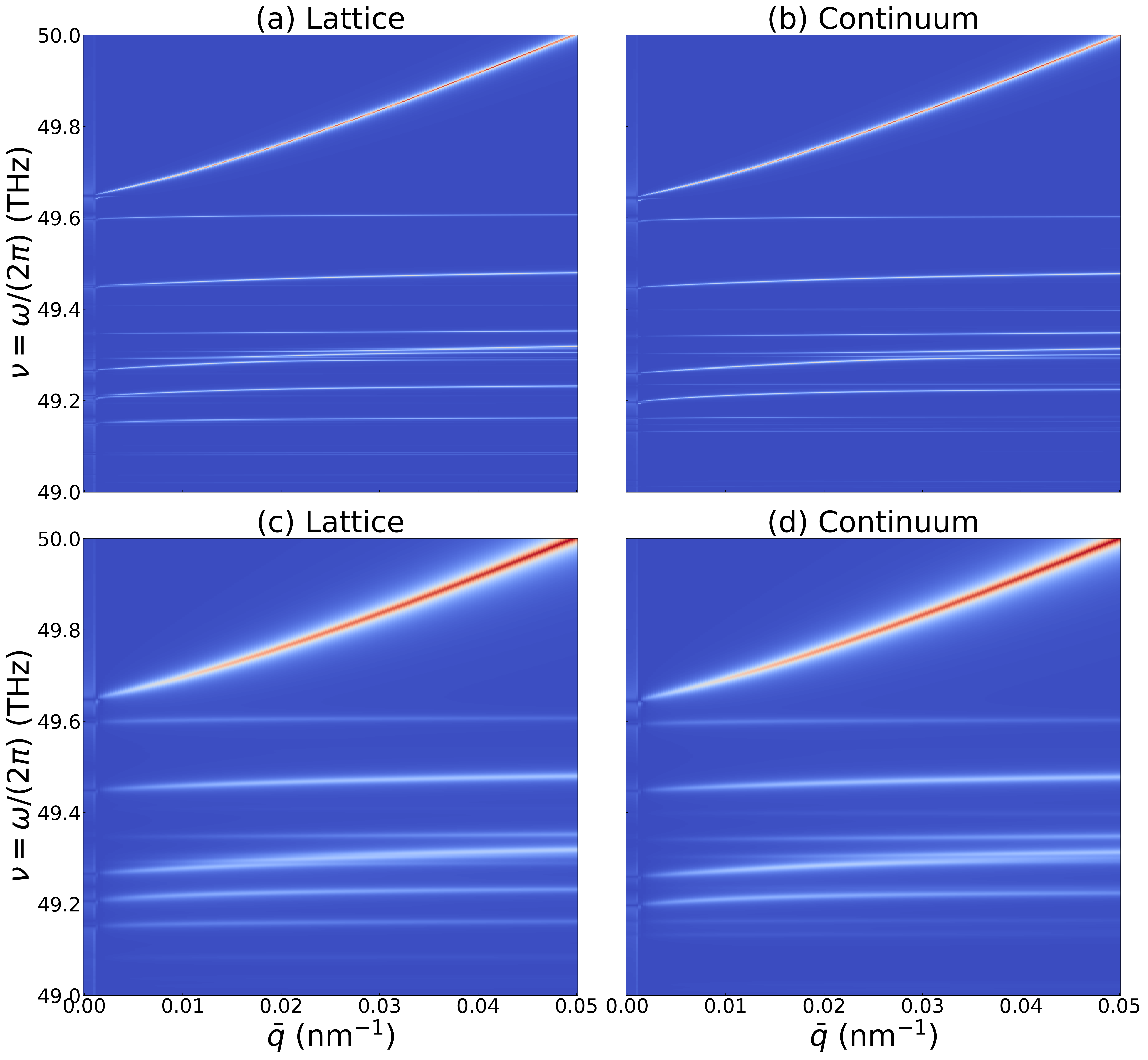}
\caption{\label{fig:fig_lattice_vs_continum} The PhP dispersion of twisted bilayer $2.65^\circ$ hBN, calculated using (a) the lattice model, $\delta/(2\pi)=0.001$ THz, (b) the continuum model, $\delta/(2\pi)=0.001$ THz, (c) the lattice model, $\delta/(2\pi)=0.01$ THz, and (d) the continuum model, $\delta/(2\pi)=0.01$ THz. The continuum model is calculated using $37$ $\bm{Q}$ vectors.}
\end{figure}

To test the accuracy of the continuum model, we plot the PhP dispersion obtained using the lattice and the continuum models in Fig. \ref{fig:fig_lattice_vs_continum} for comparison. We see that the PhP dispersion above $49.3$ THz is well captured by the continuum model. The accuracy of the continuum decreases as the frequency moves away from the critical frequency $\omega_0$, which is a feature of the $k\cdot p$ approximation. 

Our next goal is to generalize the continuum model for systems with other twisting angles. At the present stage we have only checked its accuracy for $2.65^\circ$ hBN, but we expect it to work well over a range of twisting angles. We anticipate that, similar to the BM model \cite{11pnas_BM_model}, the continuum model can only be used when the twisting angle is not too large or too small. For large angles, the continuum approximation itself is not valid. Our continuum model has a intralayer moir\'{e} potential that is much stronger than the interlayer part, so they would exhibit strong twisting angle dependence when the angle is vary small (say, $\theta<0.1^\circ$) where the corrugation effect dominates. In that case, one must be careful to tune the intralayer parameters Eq. (\ref{eq:B_11_app}). The determination of the range of applicability and the model parameters will be left for future research.

\section{More details about the moir\'{e} response function}
\subsection{Derivation in quantum case \label{append:Response_quantum_app}}
The response function used in the main text is also valid if phonons are treated quantum mechanically. We rederive it using linear response theory. The displacement operator is
\begin{align}
  \hat{\bm{u}} (\bm{r}_{Ii\alpha}) =
  \sum_{\bar{\bm{q}}b} \sqrt{\frac{\hbar}{2 M_{\alpha} \omega_{\bar{\bm{q}}b}
  N_m}} \bm{e}_{i\alpha,b}(\bar{\bm{q}}) e^{i\bar{\bm{q}} \cdot
  \bm{r}_{Ii\alpha}} (\hat{a}_{\bar{\bm{q}}b} +
  \hat{a}_{-\bar{\bm{q}}b}^{\dag}), 
\end{align}
where $\hat{a}_{\bar{\bm{q}}b}$, $\hat{a}^{\dagger}_{\bar{\bm{q}}b}$ are operators of phonon mode $\bar{\bm{q}},b$, satisfying $[\hat{a}_{\bar{\bm{q}}b},\hat{a}^{\dagger}_{\bar{\bm{q}}'b'}]=\delta_{\bar{\bm{q}}\bar{\bm{q}}'}\delta_{bb'}$.The polarization operator, defined like Eq. (\ref{eq:P_WE_moire_app}), is expanded in Fourier space
\begin{align}
  \hat{\bm{P}}(\bm{r})=\sum_{\bar{\bm{q}}\bm{Q}}e^{i(\bar{\bm{q}}+\bm{Q})\cdot\bm{r}}\hat{\bm{P}}_{\bar{\bm{q}}+\bm{Q}},
\end{align}
where $\hat{\bm{P}}_{\bar{\bm{q}}+\bm{Q}}$ can be calculated similar to Eq. (\ref{eq:Pi_qQ_cal_app}),
\begin{align}
\hat{\bm{P}}_{\bar{\bm{q}}+\bm{Q}}=\frac{e}{\Omega_m}\sum_b\bm{S}_{\bm{Q}b}(\bar{\bm{q}})\sqrt{\frac{\hbar}{2
  \omega_{\bar{\bm{q}}b} N_m}} (\hat{a}_{\bar{\bm{q}}b}+\hat{a}_{-\bar{\bm{q}}b}^{\dagger}),
\end{align}
with the form factor $\bm{S}_{\bm{Q}b}(\bar{\bm{q}})$ defined in Eq. (\ref{eq:Smat_app}).

With an electric field $\bm{E}_t(\bm{r}, t) =\sum_{\bm{Q}}\bm{E}_{\bar{\bm{q}}+\bm{Q},t}(\omega) e^{i(\bar{\bm{q}}+\bm{Q})\cdot\bm{r}-i\omega t}$, the total Hamiltonian is
\begin{align}
\hat{H}=\hat{H}_0+\hat{H}_{\text{res}},
\end{align}
where $\hat{H}_0=\sum_{\bar{\bm{q}}b}\hbar\omega_{\bar{\bm{q}}b}\hat{a}^{\dagger}_{\bar{\bm{q}}b}\hat{a}_{\bar{\bm{q}}b}$, and ($\Omega_{\text{tot}}=N_m\Omega_m$)
\begin{align}
\hat{H}_{\text{res}}=-\int d\bm{r}\hat{\bm{P}}(\bm{r})\cdot\bm{E}_t(\bm{r},t)=-\Omega_{\text{tot}}\sum_{\bm{Q}'} \hat{\bm{P}}_{-\bar{\bm{q}}-\bm{Q}'}\cdot\bm{E}_{\bar{\bm{q}}+\bm{Q}',t}(\omega) e^{-i\omega t}. 
\end{align}
Treating $\hat{H}_{\text{res}}$ as an external coupling, the induced polarization $\bm{P}_{\bar{\bm{q}}+\bm{Q}}(t) = \langle\hat{\bm{P}}_{\bar{\bm{q}}+\bm{Q}}\rangle (t) - \langle\hat{\bm{P}}_{\bar{\bm{q}}+\bm{Q}}\rangle_0$ is given by the Kubo formula as 
\begin{align}
  P_{\bar{\bm{q}}+\bm{Q},\mu}(t)=\sum_{\bm{Q}'\nu}\varepsilon_0\Pi^{\bm{Q}\bm{Q}'}_{\mu\nu}(\bar{\bm{q}},\omega)E_{\bar{\bm{q}}+\bm{Q}',\nu}(\omega)e^{-i\omega t},
\end{align}
where 
\begin{align}
\Pi^{\bm{Q}\bm{Q}'} (\bar{\bm{q}},\omega) =-\frac{N_m \Omega_m}{\varepsilon_0\hbar}\sum_{mn}\frac{[\hat{\bm{P}}_{\bar{\bm{q}}+\bm{Q}}]_{mn} [\hat{\bm{P}}^T_{-\bar{\bm{q}}-\bm{Q}'}]_{nm}}{\omega + (E_m - E_n) / \hbar + i 0^+}\frac{1}{\mathcal{Z}_0} (e^{-\beta E_m} - e^{- \beta E_n}).
\end{align}
Here $\beta=1/(k_BT)$, $\mathcal{Z}_0=\text{Tr}(e^{-\beta\hat{H}_{0}})$ is the partition function, and $[\hat{O}]_{mn}=\langle m|\hat{O}|n\rangle$ is the matrix element in the phonon Fock basis $|m\rangle,|n\rangle$ with energies $E_m$, $E_n$, respectively. Since $\hat{P}_{\bar{\bm{q}}+\bm{Q}}\propto\hat{a}_{\bar{\bm{q}}b}+\hat{a}_{-\bar{\bm{q}}b}^{\dag}$, in the summation only the following terms survive
\begin{align}
 |m\rangle=N_{n, \bar{\bm{q}}b}^{-1/2}\hat{a}_{\bar{\bm{q}}b}|n\rangle \quad\text{or}\quad |n\rangle=N_{m,-\bar{\bm{q}}b}^{-1/2}\hat{a}_{-\bar{\bm{q}}b}|m\rangle,
\end{align}
where $N_{n,\bar{\bm{q}}b}$ is the multiplicity of the $\bar{\bm{q}},b$ phonon in state $|n\rangle$. The two cases give $E_n-E_m=\hbar\omega_{\bar{\bm{q}}b}$ and $E_m - E_n = \hbar \omega_{-\bar{\bm{q}}b}$, respectively. Using Eq. (\ref{eq:time_revsersal_app}) and the bosonic statistics 
\begin{align}
\mathcal{Z}_0^{-1}\sum_ne^{-\beta E_n}N_{n,\bar{\bm{q}}b}=(e^{\beta h\omega_{\bar{\bm{q}}b}}-1)^{-1},
\end{align}
the calculation follows
\begin{align}
\Pi^{\bm{Q}\bm{Q}'} (\bar{\bm{q}}, \omega)=& - \frac{e^2}{\varepsilon_0\Omega_m}
  \sum_{m n} \sum_{bb'}\frac{\bm{S}_{\bm{Q}b}(\bar{\bm{q}})}{\sqrt{2\omega_{\bar{\bm{q}}b}}}\frac{\bm{S}^{\dagger}_{\bm{Q}'b'}(\bar{\bm{q}})}{\sqrt{2
  \omega_{\bar{\bm{q}}b'}}}\frac{[\hat{a}_{\bar{\bm{q}}b} +
  \hat{a}_{-\bar{\bm{q}}b}^{\dag}]_{m n} [\hat{a}_{-\bar{\bm{q}}b'} +
  \hat{a}_{\bar{\bm{q}}b'}^{\dag}]_{n m}}{\omega + (E_m - E_n) / \hbar + i
  0^+} \frac{1}{\mathcal{Z}_0} (e^{- \beta E_m} - e^{- \beta E_n})\nonumber \\
  =& \frac{e^2}{\varepsilon_0\Omega_m} \sum_b
  \frac{\bm{S}_{\bm{Q}b}(\bar{\bm{q}})\bm{S}^{\dagger}_{\bm{Q}'b}(\bar{\bm{q}})}{2\omega_{\bar{\bm{q}}b}}\left(\frac{1}{\omega-\omega_{\bar{\bm{q}}b}+i0^+}-\frac{1}{\omega+\omega_{\bar{\bm{q}}b}+i0^+}\right)(1 - e^{\beta \hbar \omega_{\bar{\bm{q}}b}}) \sum_n \frac{N_{n,
  \bar{\bm{q}}b}}{\mathcal{Z}_0} e^{- \beta E_n}\nonumber \\
  =& \frac{e^2}{\varepsilon_0\Omega_m} \sum_b \frac{\bm{S}_{\bm{Q}b}(\bar{\bm{q}})\bm{S}^{\dagger}_{\bm{Q}'b}(\bar{\bm{q}})}{\omega_{\bar{\bm{q}}b}^2-\omega^2-i\omega 0^+}.
\end{align}
In Ref. \cite{18a_linear_response_Pi} the authors derived a moir\'{e}-less version of the formula above in the $T=0$ limit. The derivation here indicates that the expression actually is \textit{temperature-independent}. 

\subsection{Non-locality and inhomogeneity \label{append:nonlocality_app}}
The moir\'{e} polar system realizes the spatially non-local response, in the sense that
\begin{align}
\bm{P}(\bm{r},t)=\int d\bm{r}'dt'\varepsilon_0\bm{\Pi}(\bm{r},\bm{r}',t-t')\bm{E}(\bm{r}',t').
\end{align}
Here the moir\'{e} response function, defined as
\begin{align}
\bm{\Pi}(\bm{r},\bm{r}',t)=\frac{1}{2\pi\Omega_{\text{tot}}}\int d\omega\sum_{\bar{\bm{q}}\bm{Q}\bm{Q}'}\bm{\Pi}^{\bm{Q}\bm{Q}'}(\bar{\bm{q}},\omega)e^{i(\bar{\bm{q}}+\bm{Q})\cdot\bm{r}-i(\bar{\bm{q}}+\bm{Q}')\cdot\bm{r}'}e^{-i\omega t},
\end{align}
is invariant under translations with moir\'{e} period (not the atomic cell period)
\begin{align} 
\bm{\Pi}(\bm{r},\bm{r}',t)=\bm{\Pi}(\bm{r}+\bm{L}_I,\bm{r}'+\bm{L}_I,t). \label{eq:periodicity_Pi}
\end{align}
For a general vector $\bm{a}$ that is incommensurate with the moir\'{e} lattice, the non-locality indicates $\bm{\Pi}(\bm{r},\bm{r}',t)\neq\bm{\Pi}(\bm{r}+\bm{a},\bm{r}'+\bm{a},t)$, which is different from the moir\'{e}-less case. By transforming the above formula into frequency space and setting $\bm{r}'=\bm{r}$, we obtain the local susceptibility Eq. (\ref{eq:local_Pi}) discussed in the main text.

\subsection{Symmetry properties}
For simplicity let us consider the non-degenerate case, i.e., $\omega_{\bar{\bm{q}}b}\neq\omega_{\bar{\bm{q}}b'}$ when $b\neq b'$. The time reversal requires
\begin{align}
\omega_{-\bar{\bm{q}},b}=\omega_{\bar{\bm{q}}b}, \quad \bm{e}_{i\alpha,b}(-\bar{\bm{q}})=\bm{e}^*_{i\alpha,b}(\bar{\bm{q}}),
\end{align}
while for a point group rotation $g$ of the system, it requires
\begin{align}
\omega_{g\bar{\bm{q}},b}=\omega_{\bar{\bm{q}}b},\quad g\bm{e}_{g^{-1}(i\alpha),b}(\bar{\bm{q}})=\bm{e}_{i\alpha,b}(g\bar{\bm{q}}).
\end{align}
These give the following constraints on the $S$ matrix
\begin{align}
\bm{S}_{\bm{Q}b}(\bar{\bm{q}})=&\left[\bm{S}_{-\bm{Q}b}(-\bar{\bm{q}})\right]^*, \label{eq:time_revsersal_app} \\
\bm{S}_{\bm{Q}b}(\bar{\bm{q}})=&g^{-1}\bm{S}_{g\bm{Q},b}(g\bar{\bm{q}}).
\end{align}
As a result, the response function satisfies
\begin{align}
&\Pi^{\bm{Q}\bm{Q}'}(\bar{\bm{q}},\omega)=[\Pi^{-\bm{Q},-\bm{Q}'}(-\bar{\bm{q}},-\omega)]^*,\label{eq:T_on_Pi_app}\\
&\Pi^{\bm{Q}\bm{Q}'}(\bar{\bm{q}},\omega)=g^{-1}\Pi^{g\bm{Q},g\bm{Q}'}(g\bar{\bm{q}},\omega)g.\label{eq:g_on_Pi_app}
\end{align}
The last identity also holds if there exists degeneracy [in this case we have $g\bm{S}_{\bm{Q}b_j}(\bar{\bm{q}})=\sum_{j'}\bm{S}_{g\bm{Q},b_{j'}}(g\bar{\bm{q}})U_{j'j}^g(\bar{\bm{q}})$ instead, where $b_{j(j')}$ runs over the degenerate subspace, and the matrix $U^g(\bar{\bm{q}})$ is unitary]. Besides, the form of Eq. (\ref{eq:Pi_result_app}) itself has an additional property
\begin{align}
\Pi^{\bm{Q}\bm{Q}'}(\bar{\bm{q}},\omega)=[\Pi^{\bm{Q}'\bm{Q}}(\bar{\bm{q}},-\omega)]^{\dagger}. \label{eq:Pi_dag_property}
\end{align}
If $\bm{E}(\bm{r},t)$ is an eigenmode, then so is $\bm{E}^*(\bm{r},t)$, as implied by Eq. (\ref{eq:T_on_Pi_app}). This guarantees that the eigenfields can always taken to be real. On the other hand, Eq. (\ref{eq:g_on_Pi_app}) indicates that the rotated field $g\bm{E}(g^{-1}\bm{r},t)$ is also an eigen solution with the same dispersion.

We now examine the symmetry properties of the local response function $\Pi(\bm{r},\bm{r},\omega)$, which is defined as
\begin{align}
\Pi(\bm{r},\bm{r},\omega)=\frac{1}{\Omega_{\text{tot}}}\sum_{\bar{\bm{q}}\bm{Q}\bm{Q}'}\Pi^{\bm{Q}\bm{Q}'}(\bar{\bm{q}},\omega)e^{i(\bm{Q}-\bm{Q}')\cdot\bm{r}}. \label{eq:local_Pi}
\end{align}
First, Eq. (\ref{eq:Pi_dag_property}) and the time reversal symmetry Eq. (\ref{eq:T_on_Pi_app}) lead to $\Pi^{\bm{Q}\bm{Q}'}(\bar{\bm{q}},\omega)=[\Pi^{-\bm{Q}',-\bm{Q}}(-\bar{\bm{q}},\omega)]^T$, so
\begin{align}
\Pi(\bm{r},\bm{r},\omega)=\frac{1}{\Omega_{\text{tot}}}\sum_{\bar{\bm{q}}\bm{Q}\bm{Q}'}[\Pi^{-\bm{Q}',-\bm{Q}}(-\bar{\bm{q}},\omega)]^Te^{i(\bm{Q}-\bm{Q}')\cdot\bm{r}}=\Pi^T(\bm{r},\bm{r},\omega), \label{eq:local_Pi_T}
\end{align}
i.e., the local response matrix is symmetric. Besides, using the property Eq. (\ref{eq:g_on_Pi_app}), we have
\begin{align}
\begin{split}
g\Pi(\bm{r},\bm{r},\omega)g^{-1}=&\frac{1}{\Omega_{\text{tot}}}\sum_{\bar{\bm{q}}\bm{Q}\bm{Q}'}g\Pi^{\bm{Q}\bm{Q}}(\bar{\bm{q}},\omega)g^{-1}e^{i(\bm{Q}-\bm{Q}')\cdot\bm{r}}=\frac{1}{\Omega_{\text{tot}}}\sum_{\bar{\bm{q}}\bm{Q}\bm{Q}'}\Pi^{g\bm{Q},g\bm{Q}}(g\bar{\bm{q}},\omega)e^{ig(\bm{Q}-\bm{Q}')\cdot g\bm{r}}\\
=&\frac{1}{\Omega_{\text{tot}}}\sum_{\bar{\bm{q}}\bm{Q}\bm{Q}'}\Pi^{\bm{Q}\bm{Q}}(\bar{\bm{q}},\omega)e^{i(\bm{Q}-\bm{Q}')\cdot g\bm{r}}=\Pi(g\bm{r},g\bm{r},\omega).
\end{split}
\end{align}
For a $g$-invariant point $\bm{r}_0$, i.e., there exists a moir\'{e} lattice vector $\bm{L}$ so that $g\bm{r}_0=\bm{r}_0+\bm{L}$. Using the periodic property Eq. (\ref{eq:periodicity_Pi}), we get $g\Pi(\bm{r}_0,\bm{r}_0,\omega)g^{-1}=\Pi(\bm{r}_0,\bm{r}_0,\omega)$. In other words, the local response matrix at a $g$-invariant point commutes with $g$.

For both twisted bilayer hBN and MoTe$_2$, $g=C_{3z}$ is a lattice symmetry. Now consider $\bm{r}=\bm{r}_{AA}$ and $\bm{r}_{AB}$. Both are $C_{3z}$-invariant points: $C_{3z}\bm{r}_{AA}=\bm{r}_{AA}$ and $C_{3z}\bm{r}_{AB}=\bm{r}_{AB}-\bm{L}_1^m$ [Fig. \ref{fig:fig_setup} (a)]. Therefore, we immediately find $C_{3z}\Pi^{AA(AB)}(\omega)C_{3z}^{-1}=\Pi^{AA(AB)}(\omega)$, where $\Pi^{AA(AB)}(\omega)=\Pi(\bm{r}_{AA(AB)},\bm{r}_{AA(AB)},\omega)$. On the other hand, due to Eq. (\ref{eq:local_Pi_T}), $\Pi^{AA(AB)}$ should be symmetric. As a result, $\Pi^{AA(AB)}$ must be diagonal and, in fact, proportional to the identity matrix:
\begin{align}
\Pi^{AA(AB)}(\omega)=\Pi^{AA(AB)}_{xx}(\omega)I_{2\times 2}.
\end{align}

The above discussion can be generalized. As long as the lattice has an in-plane rotation symmetry $g\neq C_{2z}$, the local response matrix at a $g$-invariant point is proportional to identity. This is because all $2\times 2$ matrices commuting with a rotation $g\neq C_{2z}$ take the form $AI_{2\times2}+B\sigma_y$ ($\sigma_y$ is the Pauli matrix, $A$, $B$ are complex numbers), and the symmetric property requires $B=0$. If the lattice has no in-plane rotation symmetry, or has only $C_{2z}$ symmetry (which is absent in polar systems, as otherwise the polarization in Eq. (\ref{eq:P_WE_moire_app}) would vanish), then at any point the local response matrix is in general not proportional to identity.

\subsection{Representation in moir\'{e}-less basis \label{appendix:monolayer_basis_app}}
In the presence of the moir\'{e} potential, the eigenvector $\bm{e}_{i\alpha,b}(\bar{\bm{q}})$ (with eigenfrequency $\omega_{\bar{\bm{q}}b}$) is related to the (folded) moir\'{e}-less eigenvectors $\bm{e}_{\alpha,la}(\bar{\bm{q}}+\bm{Q})$ (with frequency $\omega^0_{\bar{\bm{q}}+\bm{Q},la}$) by Eq. (\ref{eq:band_transform_app}) \cite{23prb_TAPW,25prb_moire_phonon_tbg}
\begin{align}
\bm{e}_{i\alpha,b}(\bar{\bm{q}})=\sum_{\bm{Q}la}\frac{e^{i\bm{Q}\cdot{(\bm{R}_i+\bm{\tau}_\alpha)}}}{\sqrt{N_a}}\bm{e}_{\alpha,la}(\bar{\bm{q}}+\bm{Q})U_{\bm{Q}la,b}(\bar{\bm{q}}).
\end{align}
The matrix $U$ is determined by the moir\'{e} potential (dynamical matrix) $D^m(\bar{\bm{q}})$, satisfying $U^{\dagger}U=UU^{\dagger}=I$ and $U_{-\bm{Q}la,b}(-\bar{\bm{q}})=U^*_{\bm{Q}la,b}(\bar{\bm{q}})$. Using the identity \cite{25prb_moire_phonon_tbg}
\begin{align}
\frac{1}{N_a}\sum_{i}e^{i\bm{Q}\cdot\bm{R}_i}=\delta_{\bm{Q}\bm{0}},
\end{align}
the moir\'{e} $S$ matrix Eq. (\ref{eq:Smat_app}) can be related to the moir\'{e}-less one Eq. (\ref{eq:Smat_moireless_app}) through
\begin{align}
\bm{S}_{\bm{Q}b}(\bar{\bm{q}})=\sqrt{N_a}\sum_a\bm{S}_{la}(\bar{\bm{q}}+\bm{Q})U_{\bm{Q}la,b}(\bar{\bm{q}}).
\end{align}
So from Eq. (\ref{eq:Pi_result_app}) we get
\begin{align}
\varepsilon_0 \bm{\Pi}^{\bm{Q}\bm{Q}'} (\bar{\bm{q}}, \omega)
=\frac{e^2}{\Omega_0}\sum_{b}\sum_{la}\sum_{l'a'}\frac{\bm{S}_{la}(\bar{\bm{q}}+\bm{Q})\bm{S}^{\dagger}_{l'a'}(\bar{\bm{q}}+\bm{Q}')}{\omega_{\bar{\bm{q}}b}^2-\omega^2}U_{\bm{Q}la,b}(\bar{\bm{q}})U^*_{\bm{Q}'l'a',b}(\bar{\bm{q}}),
\end{align}
which is equivalent to Eq. (\ref{eq:Pi_moire_continuum_app}) if one approximates $\bm{S}_{la}(\bar{\bm{q}}+\bm{Q})\approx\bm{S}_{la}(\bm{0})$. In the moir\'{e}-less case where $\bm{Q}$ is a good quantum number, $U_{\bm{Q}la,b}=\delta_{\bm{Q}la,b}$, and $\Pi(\omega)$ reduces to the diagonal moir\'{e}-less result of Eq. (\ref{eq:Pi_result_moireless_app}). We see that the off-diagonal elements of $\Pi^{\bm{Q}\bm{Q}'}$ arise from the off-diagonal elements of $U$. From perturbation theory, their strength is proportional to the moir\'{e} potential  ($\bm{Q}la\neq\bm{Q}'l'a'$, $\delta\mathcal{D}$ is the moir\'{e} potential):
\begin{align}
U_{\bm{Q}la,\bm{Q}'l'a'}(\bar{\bm{q}})\sim \frac{\bm{e}^{\dagger}_{la}(\bar{\bm{q}}+\bm{Q})\delta\mathcal{D}(\bar{\bm{q}})\bm{e}_{l'a'}(\bar{\bm{q}}+\bm{Q}')}{\omega_{\bar{\bm{q}}+\bm{Q}',l'a'}^2-\omega_{\bar{\bm{q}}+\bm{Q},la}^2}.
\end{align}

\section{Interatomic force constants \label{sec_IFC_app}}
We use the frozen phonon method to compute the interatomic force constants (FC), defined as 
\begin{align}
\Phi_{i\alpha\mu,j\beta\nu}(\bm{r}_{Ii\alpha}-\bm{r}_{Jj\beta})=\frac{\partial^2 U}{\partial u_{\mu}(\bm{r}_{Ii\alpha})\partial u_{\nu}(\bm{r}_{Jj\beta})}.
\end{align}
The potential energy $U = U_{\text{intra}}+ U_{\text{inter}}$ includes contributions from intralayer and interlayer atomic interactions. Specifically, for hBN, $U_{\text{intra}}$ is modeled by the Tersoff potential \cite{88prb_new_empirical}, and $U_{\text{inter}}$ is modeled by the registry-dependent interlayer potential \cite{05prb_registry_graphitic} tailored for the twisted bilayer hBN \cite{18nl_nanoserpents,19nl_mechanical}. For MoTe$_2$, we use Stillinger-Weber potential \cite{85prb_silicon} parameterized by Jiang \cite{15nt_SWP_MoS2} to model the intralayer interactions. For the interlayer interactions, parameters fitted for TMD systems \cite{21nl_anisotropic_interlayer, 23pca_anisotropic_interlayer} are used. 

Before computing the FCs, the conjugate-gradient and \textit{fire} minimization algorithms are sequentially performed using the large-scale atomic/molecular massively parallel simulator (LAMMPS) \cite{22cpc_lammps} to optimize the simulation cell and atomic positions. We have intentionally avoided performing the non-analytical correction on the dynamical matrix, because otherwise it results in a double counting of the Coulomb force. We note that different FCs computed using different force fields could indeed quantitatively influence the polariton dispersion, since the moir\'{e} physics occurs on tiny energy scales. Nevertheless, the qualitative moir\'{e} structure of the PhP dispersion should be robust and will not be altered by quantitative differences.

\end{document}